\documentclass[twocolumn,twoside,pra,superscriptaddress]{revtex4}
\usepackage[utf8]{inputenc}

\usepackage[pdftex]{graphicx,color}
\usepackage{amsmath}
\usepackage{amssymb}
\usepackage{epsfig}
\usepackage{amsthm}
\usepackage{mathtools}
\usepackage{tikz}
\usepackage{wrapfig}
\usepackage{comment}
\usepackage{simpler-wick}
\usepackage{simplewick}
\usepackage{latexsym,revsymb}
\usepackage[unicode=true,pdfusetitle,
bookmarks=true,bookmarksnumbered=false,bookmarksopen=false,
breaklinks=false,pdfborder={0 0 0},backref=false,colorlinks=false]
{hyperref}

\newcommand{\ket}[1]{| #1 \rangle}

\newcommand{\vertiii}[1]{{\left\vert\kern-0.25ex\left\vert\kern-0.25ex\left\vert #1 
    \right\vert\kern-0.25ex\right\vert\kern-0.25ex\right\vert}}

\makeatletter
\newtheorem*{rep@theorem}{\rep@title}
\newcommand{\newreptheorem}[2]{%
\newenvironment{rep#1}[1]{%
 \def\rep@title{#2 \ref{##1}}%
 \begin{rep@theorem}}%
 {\end{rep@theorem}}}
\makeatother

\newreptheorem{theorem}{Theorem}

\newreptheorem{corollary}{Corollary}

\newreptheorem{lemma}{Lemma}

\begin{document}
\title{Discretizing quantum field theories for quantum simulation}
\author{Terry Farrelly}
\email{farreltc@tcd.ie}
\affiliation{Institut f{\"u}r Theoretische Physik, Leibniz Universit{\"a}t Hannover, Germany}
\affiliation{ARC Centre for Engineered Quantum Systems, School of Mathematics and Physics, University of Queensland, Brisbane, QLD 4072, Australia}
\author{Julien Streich}
\email{julien-streich@gmx.de}
\affiliation{Institut f{\"u}r Theoretische Physik, Leibniz Universit{\"a}t Hannover, Germany}

\begin{abstract}
To date, all proposed quantum algorithms for simulating quantum field theory (QFT) simulate (continuous-time) Hamiltonian lattice QFT as a stepping stone.  Two overlooked issues are how large we can take the timestep in these simulations while getting the right physics and whether we can go beyond the standard recipe that relies on Hamiltonian lattice QFT.  The first issue is crucial in practice for, e.g., trapped-ion experiments which actually have a lower bound on the possible ratio of timestep to lattice spacing.  To this end, we show that a timestep equal to or going to zero faster than the spatial lattice spacing is necessary for quantum simulations of QFT, but far more importantly a timestep equal to the lattice spacing is actually \emph{sufficient}.  To do this, first for $\phi^4$ theory, we give a quantum circuit exactly equivalent to the real-time path integral from the discrete-time Lagrangian formulation of lattice QFT.  Next we give another circuit with no lattice QFT analogue, but, by using Feynman rules applied to the circuit, we see that it also reproduces the correct continuum behaviour.  Finally, we look at non-abelian gauge fields, showing that the discrete-time lattice QFT path-integral is exactly equivalent to a finite-depth local circuit.  All of these circuits have an analogue of a lightcone on the lattice and therefore are examples of quantum cellular automata.  Aside from the potential practical benefit of these circuits, this all suggests that the path-integral approach to lattice QFT need not be overlooked in quantum simulations of physics and has a simple quantum information interpretation.
\end{abstract}

\maketitle

\section{Introduction}
Calculations in quantum field theory become particularly difficult when perturbative methods no longer work (usually when interactions are strong).  To get around this, lattice QFT was introduced \cite{74Wilson,Creutz83,DeGrandDeTar,Maas2017} to regulate QFT and allow classical computers to perform calculations.  Lattice quantum chromodynamics has been particularly successful in calculating particle mass ratios \cite{08Duerr} and studying phases of gauge theories \cite{79Creutz,08AFN}.  Unfortunately, dynamical simulations of lattice QFT suffer from the same problem that most classical simulations of quantum many-body systems do:\ the memory required grows exponentially  with the system size.  To remedy this, quantum algorithms for simulations of some QFTs were introduced \cite{06BY,JLP12,14aJLP,JLP14,15BRS,15MPS,JKL18,ABL19,19MGJ}, and it is conceivable that these will find application in the near term on relatively small quantum computers \cite{MHM17,19YDM,Preskill2018}, with pioneering experiments already being performed \cite{MMS16,BBC19}.
\begin{figure}[!ht]
{\centering
\resizebox{8.6cm}{!}{\includegraphics{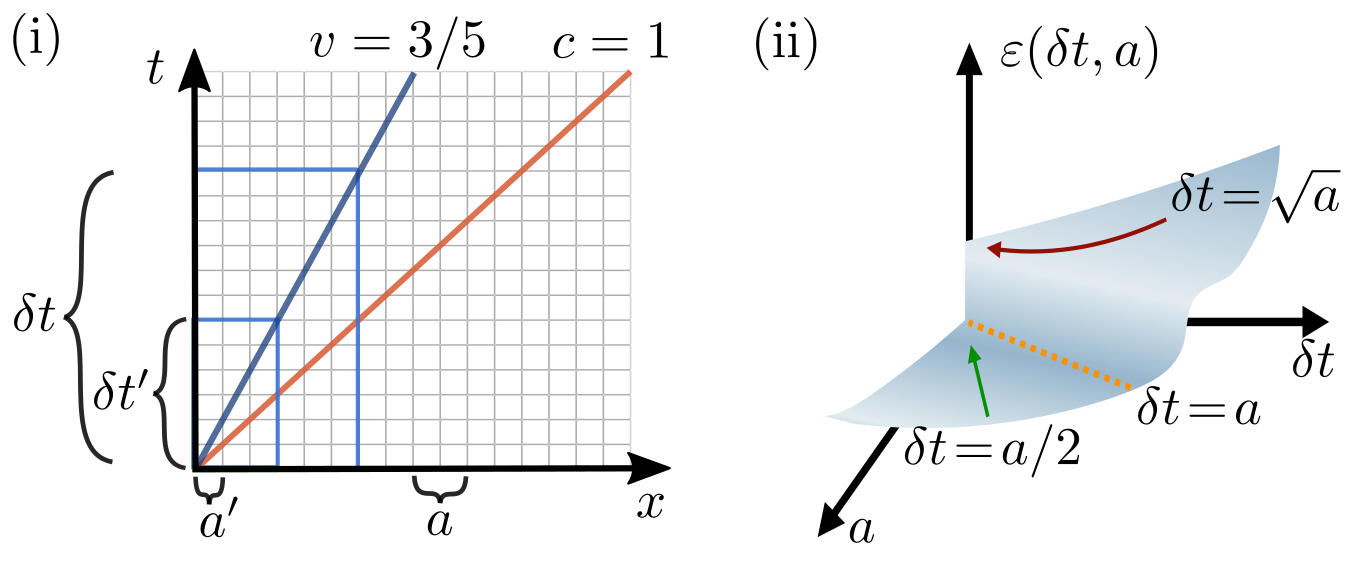}} \caption[Intuition]{(i) If $\delta t/a$ is too big the simulation cannot reproduce the correct physics at high velocities regardless of how small $a$ is.  E.g., for a nearest-neighbour circuit with $\delta t = 10a$.  As we fine grain to $a^{\prime}=a/2$, the maximum speed in the circuit is still $v=3/5$ in units with $c=1$.  (ii) Intuitive sketch of the error in a QFT simulation $\varepsilon(\delta t,a)$ as a function of $a$ and $\delta t$.  Simulations with $\delta t/a$ going to zero too slowly will never converge to the right physics, but those with $\delta t/a\leq 1$ as $a\rightarrow 0$ can. \label{fig:main}}
}
\end{figure}

All previously proposed quantum algorithms for QFT (e.g., \cite{JLP12,JLP14}) approximate continuous-time lattice Hamiltonian dynamics by a quantum circuit (via a Suzuki-Trotter decomposition) with a timestep $\delta t$.  Then by taking $\delta t$ sufficiently small, one can approximate the dynamics of the lattice QFT (with a fixed spatial lattice spacing $a$) to the desired accuracy.  One repeats this procedure for smaller and smaller $a$ to extrapolate to the continuum limit.  In contrast, here we argue that choosing a quantum circuit with $\delta t= a$ is sufficient to simulate QFTs.  This is vital for experiments with technical limitations leading to a \emph{minimal allowed} $\delta t/a$.  For example, this is actually necessary to avoid unwanted spin-motion coupling when applying M{\o}lmer-S{\o}rensen gates in trapped ions \cite{MHM17a}, a consequence of which is that the sub-timesteps (that form parts of one full Trotter step) that fix $\delta t/a$ cannot be too small.  Furthermore, in current experiments it is desirable to minimize the number of gates to reduce gate errors.  In fact, the scaling $\delta t= a$ (in units with $c=1$) is essentially the best we could hope for for circuits with fixed interaction range (e.g., nearest neighbour or next-nearest neighbour).  Any ``better'' scaling, say $\delta t\propto a^{1/2}$, would mean that the lightcone of the circuit will be narrower than the physical lightcone as $a\rightarrow 0$, and the circuit will fail to simulate all of the physics.  This is explained intuitively in figure \ref{fig:main} and is analogous to the CFL condition for hyperbolic PDEs \cite{CFL28}.

We start with quantum-circuit discretizations of interacting scalar field theory that have some appealing properties:\ the number of local unitary gates is proportional to the spacetime volume, and there is a strict upper bound on the speed of information propagation.  There is more than one possible circuit with these properties, but our first example is very natural because it is equivalent to the discrete-time path integral.  We introduce another circuit, not arising from a Suzuki-Trotter decomposition or path integral, which has the right continuum limit in the free case but suffers from a bosonic analogue of fermion doubling, which can be remedied with a slightly modified interaction.  This illustrates an important point:\ it is possible (and may be pragmatic) to use circuits not arising from conventional lattice QFT but still giving the right continuum limit.  Finally, we look at pure non-abelian gauge theory on the lattice, where we see that the discrete-time path integral amplitude is exactly equal to a sequence of finite-depth local circuits with timestep $\delta t= a$.  These results, aside from having practical significance, also give a quantum information perspective on the discrete-time path integrals for bosonic lattice QFT:\ we show they are equivalent to simple quantum circuits.
 
These quantum-circuits are also quantum cellular automata, which are discrete spacetime quantum systems where information propagates at a bounded speed \cite{19Farrelly,19Arrighi}.  Having strictly bounded propagation speeds is actually impossible for continuous-time lattice systems with local Hamiltonians \cite{Farrelly15}.  There have been proposals for discretizing QFT via quantum cellular automata in the past, but most corresponded to free QFTs \cite{Bisio2012,Bisio2014,BDP15a,DAriano2016,Bisio2016} or integrable QFTs in one space dimension \cite{DdV87,BDP18,ABF19}.  (Another approach discretizes fermions interacting with an abelian gauge field \cite{Yepez16b,Yepez16}, but it is not clear whether the fermion doubling problem \cite{NN81} can be avoided.)

\section{Scalar fields on a lattice}
To start, consider the continuous-time Hamiltonian version of lattice QFT \cite{KS75} for $\phi^4$ theory \cite{PeskinSchroeder}.  Before now, this has always been the formalism of choice for quantum simulations of QFT.  Nevertheless, we will argue that the discrete-time Lagrangian formalism can be more natural for quantum simulations.  Of course, in the continuum limit, all formalisms agree.

We discretize space with coordinates given by vectors of integers $\mathbf{n}\in \mathbb{Z}^d$ and spatial lattice spacing $a$.  At each lattice site there are field operators $\phi_{\mathbf{n}}$ and $\pi_{\mathbf{n}}$ that satisfy $\left[\phi_{\mathbf{m}},\pi_{\mathbf{n}}\right]=\frac{i}{a^d}\delta_{\mathbf{m},\mathbf{n}}$, as well as $[\phi_{\mathbf{m}},\phi_{\mathbf{n}}]=[\pi_{\mathbf{m}},\pi_{\mathbf{n}}]=0$ for all $\mathbf{m}$ and $\mathbf{n}$.  The factor of $1/a^d$ ensures that we recover the continuum commutation relations as $a\rightarrow 0$.
We also write $\mathbf{x} = \mathbf{n}a$ and $t = \tau \delta t$, where $\tau\in\mathbb{Z}$ is the discrete time coordinate.  For simplicity, we take the spatial lattice to be infinite, but everything extends to finite lattices (with periodic boundary conditions), which is necessary for simulations in practice.

The continuous-time dynamics is described by a Hamiltonian, with one possible choice being
\begin{equation}\label{eq:lattHam}
 H_{\mathrm{latt}}=a^d\sum_{\mathbf{n}\in \mathbb{Z}}\left[\frac{1}{2}\pi_{\mathbf{n}}^2+\frac{1}{2}(\nabla_a \phi_{\mathbf{n}})^2+\frac{m^2}{2}\phi_{\mathbf{n}}^2+\frac{\lambda}{4!}\phi_{\mathbf{n}}^4\right],
\end{equation}
with the simplest choice for the discrete gradient:
\begin{equation}
 (\nabla_a \phi_{\mathbf{n}})^2=\sum_{\mathbf{e}\in\mathcal{N}}\left(\frac{\phi_{\mathbf{n}+\mathbf{e}}-\phi_{\mathbf{n}}}{a}\right)^2,
\end{equation}
where $\mathcal{N}=\{(1,0,...,0),...,(0,...,0,1)\}$ are lattice basis vectors.  It is useful to write $H_{\mathrm{latt}}= H_P + H_X$, where $H_P$ consists of all terms with $\pi_n$ and $H_X$ includes all terms with $\phi_n$.

Although spacetime is now discrete, field operators still describe continuous degrees of freedom.  On quantum computers made of qubits (as opposed to continuous variables) the on-site fields need to be discretized and truncated, which is not as costly as one might imagine \cite{KS19}.  An intriguing alternative is to use continuous-variable quantum computers for, e.g., $\phi^4$ theory \cite{15MPS}.  Nevertheless, how the fields are discretized is not our concern here.  Instead, we focus on the dynamics of the fields with the understanding that a representation of the fields tailored to the specific architecture can be chosen when necessary.

\section{Strang-split circuit for scalar field theory}
Our first circuit actually comes from Strang splitting (a special case of a Suzuki-Trotter decomposition \cite{90Suzuki}) the evolution operator of the Hamiltonian lattice QFT, $\exp(-iH_{\mathrm{latt}}\delta t)$.  (The other circuits we will consider for $\phi^4$ theory and non-abelian gauge fields do not come from such a decomposition of a lattice Hamiltonian.)  This seems like business as usual for a quantum simulation of a QFT, but we depart from conventional quantum algorithms by fixing the timestep $\delta t= a$.  This is the key point, as we will still get the correct continuum limit with this circuit in spite of this restriction.  The circuit evolving the system over one timestep is 
\begin{equation}\label{eq:Strang-ver}
 U=e^{-iH_X\delta t/2}e^{-iH_P\delta t}e^{-iH_X\delta t/2}.
\end{equation}
Each of $e^{-iH_X\delta t/2}$ or $e^{-iH_P\delta t}$ is a product of commuting local unitaries, so $U$ is a finite-depth local unitary circuit.

An advantage of this circuit is that we need not think of it as approximating the continuous-time lattice Hamiltonian dynamics, but rather as exactly simulating the path integral in the discrete-time Lagrangian formulation of lattice QFT.  This follows because
\begin{equation}\label{eq:phi-path-strang}
 \langle \varphi_{\mathrm{f}}|U^{\tau}|\varphi_{\mathrm{i}}\rangle = \int \mathcal{D}(\varphi)e^{iS(\varphi)},
\end{equation}
where $|\varphi_{\mathrm{i}}\rangle$ and $|\varphi_{\mathrm{f}}\rangle$ are initial and final states at time $0$ and time $\tau$ respectively.  The action is
\begin{equation}
S(\varphi) = \sum_{\nu=0}^{\tau-1}\delta t\sum_{\mathbf{n}}a^d\mathcal{L}(\nu,\mathbf{n}),
\end{equation}
where
\begin{equation}\label{eq:Lfirst}
\begin{split}
\mathcal{L}(\nu,\mathbf{n}) & = \frac{(\varphi_{\nu+1,\mathbf{n}}-\varphi_{\nu,\mathbf{n}})^2}{2\delta t^2}-
\frac{\mathcal{V}(\varphi_{\nu+1,\mathbf{n}})+\mathcal{V}(\varphi_{\nu,\mathbf{n}})}{2},
\end{split}
\end{equation}
with $\mathcal{V}(\varphi_{\nu,\mathbf{n}})$ containing all spatial derivatives, mass term and the interaction:
\begin{equation}
\mathcal{V}(\varphi_{\nu,\mathbf{n}}) = \left[\frac{1}{2}(\nabla_a \varphi_{\nu,\mathbf{n}})^2+\frac{m^2}{2}\varphi_{\nu,\mathbf{n}}^2+\frac{\lambda}{4!}\varphi_{\nu,\mathbf{n}}^4\right],
\end{equation}
where
\begin{equation}
 (\nabla_a \varphi_{\nu,\mathbf{n}})^2=\sum_{\mathbf{e}\in\mathcal{N}}\left(\frac{\varphi_{\nu,\mathbf{n}+\mathbf{e}}-\varphi_{\nu,\mathbf{n}}}{a}\right)^2.
\end{equation}
$\varphi_{\nu,\mathbf{n}}$ is the classical scalar field at time $\nu$ at position $\mathbf{n}$.  The symmetrized form of the potential $\mathcal{V}$ in the Lagrangian in equation (\ref{eq:Lfirst}) is simply analogous to averaging a left and a right Riemann sum approximation to the continuum action.

The proof of equation (\ref{eq:phi-path-strang}) is standard, although usually given in the Euclidean setting with imaginary time (and periodic boundary conditions in time) \cite{Smit02}.  For completeness, we give the proof in appendix \ref{sec:Strang-proof}.  From the perspective of lattice QFT, $U$ is the real-time transfer matrix.  The significant point for us here is that this $U$ is \emph{exactly} a local quantum circuit.  This means that we can simulate the dynamics with the choice $\delta t = a$ and get the correct continuum limit, which not only has a practical advantage but also gives a quantum information perspective on the path integral for this model:\ it is just a quantum circuit.

The first-order Trotter decomposition of $H_{\mathrm{latt}}$, i.e., $U_{\mathrm{Trott}}=e^{-iH_P\delta t}e^{-iH_X\delta t}$, gives a simpler Lagrangian: $\mathcal{L}(\nu,\mathbf{n})  = \frac{(\varphi_{\nu+1,\mathbf{n}}-\varphi_{\nu,\mathbf{n}})^2}{2\delta t^2}-
\mathcal{V}(\varphi_{\nu,\mathbf{n}})$.  But we expect the Strang-split circuit to converge faster to the continuum limit.  Another advantage to the Strang-split circuit is not only more accuracy compared to $U_{\mathrm{Trott}}$, but it only needs one additional layer of unitary gates since $U^{\tau}=e^{-iH_X\delta t/2}(U_{\mathrm{Trott}})^{\tau-1}e^{-iH_P\delta t}e^{-iH_X\delta t/2}$, i.e., $U_{\mathrm{Trott}}^{\tau}$ consists of $2\tau$ alternating layers like $e^{-iH_P\delta t}$, whereas $U^{\tau}$ consists of $2\tau+1$. 

It is important to reiterate the main point:\ quantum simulations with $\delta t = a$ are sufficient to capture the right physics.  This is because we have shown that the path integral amplitude from discrete-time Lagrangian lattice QFT \emph{is exactly equal to a local quantum circuit}, which we know has the right continuum limit as the choice $\delta t=a$ is sufficient in lattice QFT calculations to get the right physics \cite{Durr13}.

\section{Shift circuit for scalar field theory}
Let us introduce a second circuit for $\phi^4$ theory, which we call the Shift circuit.  We take $U=W_XW_PW_X$, with
\begin{equation}\label{eq:shift-dec}
\begin{split}
 W_X & = \prod_{\mathbf{n}\in \mathbb{Z}^d} \exp\!\left[ia^{d-1}\!\!\left(\frac{M}{2^{D}}\sum_{\mathbf{e}\in N}\phi_{\mathbf{n}}\phi_{\mathbf{n}+\mathbf{e}}+\frac{\lambda a^2}{4!}\phi_{\mathbf{n}}^4\!\right)\!\right]\\
 W_P & = \prod_{\mathbf{n}\in \mathbb{Z}^d} \exp\!\left[-\frac{i\pi a^{d-1}}{4}(\phi_{\mathbf{n}}^2+a^2\pi_{\mathbf{n}}^2)\right].
 \end{split}
\end{equation}
Here $D=d+1$ and $M=1-m^2a^2/2$, where $m$ is the mass, and $N$ are all lattice vectors $\mathbf{e}$ with components $e_i=\pm 1$.  Again $\delta t = a$.
This circuit is a natural discretization of the continuum theory when $m=\lambda=0$ in $d=1$.  This is because, in that case, the continuum model can be solved in terms of left- and right-moving fields, which also actually works for this circuit as shown in appendix \ref{sec:Shift-motivation}.

With $\lambda =0$ the dynamics can be solved exactly, with the circuit's Feynman propagator being given by (see appendix \ref{app:Feyn_prop})
\begin{equation}
\begin{split}
 D_F(x-y)  = \frac{a^2}{2}\!\int\! \frac{\mathrm{d}^Dp}{(2\pi)^D}\frac{ie^{-ip.(x-y)}}{M\prod_{i=1}^d\cos(p_i a)-\cos[p_0 a]+i\varepsilon},
 \end{split}
 \end{equation}
where integral is over $(-\pi/a,\pi/a]^D$.  As $a\rightarrow 0$, we recover the usual Feynman propagator.

The Strang-split circuit was clearly a good regulator, but is this true for the Shift circuit?  To check, we use the Feynman rules for the circuit (derived more generally in appendix \ref{sec:Feynman-rules-discrete}) to calculate the order $\lambda$ correction to the particle mass in $D=1+1$.  This gives $\Pi_{\mathrm{Shift}}(p_{\mathrm{in}}^2)  = \frac{\lambda}{2\pi}\ln\left(\frac{1}{ma}\right)+\mathrm{finite}$ (calculated in appendix \ref{app:One_loop}), which is actually too big by a factor of two compared to other regulators, e.g., continuum QFT with a momentum cutoff $\Lambda$, which gives
\begin{equation}
\Pi_{\mathrm{cont}}(p_{\mathrm{in}}^2) =\frac{\lambda}{4\pi}\ln\left(\frac{\Lambda}{m}\right)+\mathrm{finite} 
\end{equation}
(also calculated in appendix \ref{app:One_loop} for completeness).
  The problem is that the equal-time Shift circuit propagator has a symmetry under $p_i\leftrightarrow \pi/a -p_i$.  These high-momentum modes act like low-momentum modes, contributing to Feynman diagrams like a second type of particle.  To fix this, we use the smeared interaction term
 \begin{equation}
V = a^{D}\frac{\lambda}{4!}\sum_{\mathbf{n}\in\mathbb{Z}^d}\left(\sum_{\mathbf{e}\in \mathcal{K}}w(\mathbf{e})\phi_{\mathbf{n}+\mathbf{e}}\right)^{4},
 \end{equation}
 where $\mathcal{K}$ are all vectors with components in $\{-1,0,1\}$ and $w(\mathbf{e})=v(e_1)\times...\times v(e_d)$ with $v(-1)=v(1)=1/4$ and $v(0)=1/2$.  This modifies the Feynman rules only at high energy, so that the vertex factor depends on the momentum flowing in or out.  At low energy, we get the new correct value for the diagram $\Pi_{\mathrm{Shift}}(p_{\mathrm{in}}^2) = \frac{\lambda}{4\pi}\ln\left(\frac{1}{ma}\right)+\mathrm{finite}$
 as long as $p_{\mathrm{in}}\ll 1/a$, which is calculated in appendix \ref{app:One_loop}.  So with the modified interaction we get the correct value for the first order correction to the mass because high-momentum particles are now suppressed.

\section{Local circuit for non-abelian gauge theory}
Let us turn to pure non-abelian gauge theory on the lattice, which, despite the absence of matter, is interacting.  The basic starting point here is simple:\ the path integral can be written in terms of a real-time transfer operator.  In this case, this does not come from a Trotter decomposition of the lattice Hamiltonian dynamics, which again has always been the usual approach for quantum simulations of QFTs.  The essential new point is that the transfer operator is actually \emph{exactly} a finite-depth local quantum circuit, which allows us to simulate the dynamics with a quantum circuit with $\delta t = a$.

Here the Hilbert spaces live on the links between lattice sites.  We denote lattice sites by $\mathbf{x}$ with components in $a\mathbb{Z}$, but now $\mathbf{e}\in\{(a,0,...,0),...,(0,...,0,a)\}$ are lattice basis vectors. The Hilbert space for the link $(\mathbf{x},\mathbf{x}+\mathbf{e})$ is spanned by the basis $|U_{\mathbf{x},\mathbf{e}}\rangle$, where $U_{\mathbf{x},\mathbf{e}}$ denotes an element of the defining representation of the gauge group $\mathcal{G}$, which we assume is either $\mathrm{U}(1)$ or $\mathrm{SU}(n)$.  (Non-compact groups can also be considered.)  In other words, $U_{\mathbf{x},\mathbf{e}}$ corresponds to a matrix of real numbers $U_{\mathbf{x},\mathbf{e}}^{ab}$, where $a$ and $b$ are the matrix indices, e.g., for $\mathrm{SU}(n)$, these run from $1$ to $n$, whereas for $\mathrm{U}(1)$ there is only a single element.  These basis states satisfy
\begin{align}
 \langle V_{\mathbf{x},\mathbf{e}}|U_{\mathbf{x},\mathbf{e}}\rangle & = \delta(V_{\mathbf{x},\mathbf{e}},U_{\mathbf{x},\mathbf{e}})\\
 \int \mathrm{d}U_{\mathbf{x},\mathbf{e}} |U_{\mathbf{x},\mathbf{e}}\rangle\langle U_{\mathbf{x},\mathbf{e}}| & = \openone,
\end{align}
where $\mathrm{d}U_{\mathbf{x},\mathbf{e}}$ denotes the normalized Haar measure over the group and $\int \mathrm{d}U\delta(U,V)=1$.  We also associate operators $\hat{U}_{\mathbf{x},\mathbf{e}}^{ab}$ to each link (here it is better to give operators hats) with
$\hat{U}_{\mathbf{x},\mathbf{e}}^{ab} |U_{\mathbf{x},\mathbf{e}}\rangle = U_{\mathbf{x},\mathbf{e}}^{ab} |U_{\mathbf{x},\mathbf{e}}\rangle$.
One important operator is given by the trace of spatial plaquette operators on the lattice.  These are
\begin{equation}
 \mathrm{tr}[\hat{U}_{\mathbf{p}_s}]=\hat{U}_{\mathbf{x},\mathbf{e}}^{ab}\hat{U}_{\mathbf{x}+\mathbf{e},\mathbf{f}}^{bc}\hat{U}_{\mathbf{x}+\mathbf{f},\mathbf{e}}^{\dagger cd}\hat{U}_{\mathbf{x},\mathbf{f}}^{\dagger da},
\end{equation}
where repeated matrix indices are summed, and $\mathbf{p}_s=(\mathbf{x},\mathbf{x}+\mathbf{e},\mathbf{x}+\mathbf{e}+\mathbf{f},\mathbf{x}+\mathbf{f})$ denotes a spatial plaquette.  We also define left shift operators that act on link states via
\begin{equation}
 \hat{L}_{\mathbf{x},\mathbf{e}}(V)|U_{\mathbf{x},\mathbf{e}}\rangle = |V^{\dagger}U_{\mathbf{x},\mathbf{e}}\rangle.
\end{equation}

Finally, all physical states must be invariant under spatial gauge transformations (which is Gauss' law)
\begin{equation}
 U_{\mathbf{x},\mathbf{e}}\rightarrow \Omega_{\mathbf{x}}U_{\mathbf{x},\mathbf{e}}\Omega^{\dagger}_{\mathbf{x}+\mathbf{e}} = U^{\Omega}_{\mathbf{x},\mathbf{e}},
 \end{equation}
where $\Omega_{\mathbf{x}}$ are in the gauge group.  A unitary operator implementing a gauge transformation is
$\hat{D}(\Omega)|U\rangle = | U^{\Omega}\rangle$, where $| U\rangle=\prod_{\mathbf{x},\mathbf{e}}|U_{\mathbf{x},\mathbf{e}}\rangle$ is a state of the lattice.
Gauss' law is then that any \emph{physical} state $\ket{U}$ satisfies $\hat{D}(\Omega)|U\rangle = | U\rangle$ for any $\Omega$.

Next consider the unitary evolution operator (real-time transfer operator) $\hat{T}$, given by $\hat{T}=\hat{W}_{\mathrm{el}}\hat{W}_{\mathrm{mag}}$,
where
\begin{equation}\label{eq:WmagWel}
 \begin{split}
  \hat{W}_{\mathrm{mag}} & =\exp\left(-i\frac{2}{g^2}\sum_{\mathbf{p}_s}\mathrm{Re}\left(\mathrm{tr}[\hat{U}_{\mathbf{p}_s}]\right)\right)\\
  \hat{W}_{\mathrm{el}} & =\prod_{\mathbf{x},\mathbf{e}}\int \mathrm{d}V_{\mathbf{x},\mathbf{e}} \exp\left(-i\frac{2}{ g^2}\mathrm{Re}\left(\mathrm{tr}[V_{\mathbf{x},\mathbf{e}}]\right)\right)\hat{L}_{\mathbf{x},\mathbf{e}}(V_{\mathbf{x},\mathbf{e}}),
 \end{split}
\end{equation}
where $g$ is the gauge coupling constant, and $\mathrm{Re}(\mathrm{tr}[A])=(\mathrm{tr}[A]+\mathrm{tr}[A^{\dagger}])/2$.  Note that the product over plaquettes counts each plaquette once.  The crucial point is that the terms in the exponents in the definition of $\hat{W}_{\mathrm{mag}}$ all commute since plaquette operators $\hat{U}_{\mathbf{p}_s}$ commute with each other.  As a result, both $\hat{W}_{\mathrm{mag}}$ and $\hat{W}_{\mathrm{el}}$ are finite-depth local quantum unitaries.  The depth of $\hat{W}_{\mathrm{el}}$ is one since each unitary acts on a different link, whereas the depth of $\hat{W}_{\mathrm{mag}}$ depends on the lattice dimension because plaquette operators may overlap on links.  In $d=2$ the depth is two:\ unitaries on plaquettes can be applied in a chess-board pattern.  More generally, a depth of $d!$ is enough (by using the $d=2$ method for each pair of lattice directions).

We chose this evolution operator $\hat{T}$ because it is equivalent to a discrete-time path integral with the Wilson action, i.e., given initial and final gauge field configurations obeying Gauss' law, $|U_i\rangle$ and $|U_f\rangle$,
\begin{equation}\label{eq:gauge-equiv}
 \langle U_f | \hat{T}^{\tau}|U_i\rangle=\int \mathcal{D}(U)e^{iS(U)},
\end{equation}
where the action is
\begin{equation}
S(U) = \frac{2}{g^2}\sum_{p}\mathrm{Re}\left(\mathrm{tr}[U_{p}]\right),
\end{equation}
and we have $\mathrm{tr}[U_{p}]=U_{x,e}^{ab}U_{x+e,f}^{bc}U_{x+f,e}^{* cd}U_{x,f}^{* da}$, where $p=(x,x+e,x+e+f,x+f)$ is a spacetime plaquette, with $x$ representing a spacetime coordinate, and $e$ and $f$ are spacetime lattice basis vectors.  The range of the sum is over all spacetime plaquettes on $\{0,..,\tau\}\times \mathbb{Z}^d$ except spatial plaquettes at time $\tau$.  The measure is $\mathcal{D}(U)=\prod_{x,e}\mathrm{d}U_{x,e}$.
For completeness, the derivation of equation (\ref{eq:gauge-equiv}) is given in appendix \ref{app:non-ab-circ}.

The unitaries $\hat{W}_{\mathrm{mag}}$ and $\hat{W}_{\mathrm{el}}$ are local finite-depth circuits, but their precise form will depend on how we represent the link Hilbert spaces (on either qubits or continuous variables).  Any local truncation scheme for the link degrees of freedom can be used in conjunction with the above circuit, but it is worthwhile to point out that finding a good truncation scheme is difficult.  Two possible options are to use the group theoretic methods from \cite{06BY} or quantum link models \cite{CW97} to truncate the link Hilbert spaces to represent them on qubits.  For example, for the quantum-link version of an $\mathrm{SU}(2)$ gauge theory, it is sufficient to have two qubits per link \cite{CW97}, while maintaining Gauss' law exactly.

As a final note, the unitary $\hat{W}_{\mathrm{el}}$ looks a little mysterious, but it is not as complicated as it seems.  For a start, it involves a product of the same unitary on every link, and these commute with gauge transformations.  (Any gauge theory should have the property that the dynamics commutes with gauge transformations.)  To verify that these individual unitaries are guage invariant, we first need
\begin{equation}
\begin{split}
 \hat{L}_{\mathbf{x},\mathbf{e}}(V)\hat{D}(\Omega)|U_{\mathbf{x},\mathbf{e}}\rangle & = |V^{\dagger}\Omega_{\mathbf{x}}U_{\mathbf{x},\mathbf{e}}\Omega^{\dagger}_{\mathbf{x}+\mathbf{e}}\rangle\\
 & = |\Omega_{\mathbf{x}}\left(\Omega^{\dagger}_{\mathbf{x}}V^{\dagger}\Omega_{\mathbf{x}}\right)U_{\mathbf{x},\mathbf{e}}\Omega^{\dagger}_{\mathbf{x}+\mathbf{e}}\rangle\\
 & = \hat{D}(\Omega)\hat{L}_{\mathbf{x},\mathbf{e}}\left(\Omega^{\dagger}_{\mathbf{x}}V\Omega_{\mathbf{x}}\right)|U_{\mathbf{x},\mathbf{e}}\rangle.
\end{split}
 \end{equation}
But this together with the definition of $\hat{W}_{\mathrm{el}}$ in equation (\ref{eq:WmagWel}) implies that $\hat{W}_{\mathrm{el}}$ commutes with $\hat{D}(\Omega)$.  This follows because $\mathrm{tr}[\Omega^{\dagger}_{\mathbf{x}}V_{\mathbf{x},\mathbf{e}}\Omega_{\mathbf{x}}]=\mathrm{tr}[V_{\mathbf{x},\mathbf{e}}]$ and because the integral is over the Haar measure which satisfies $\mathrm{d}V_{\mathbf{x},\mathbf{e}}=\mathrm{d}(\Omega^{\dagger}_{\mathbf{x}}V_{\mathbf{x},\mathbf{e}}\Omega_{\mathbf{x}})$.  This simplifies the possibilities for the unitary gates in $\hat{W}_{\mathrm{el}}$ on each link (or approximations to them), since they must be functions of the invariants of the gauge group on the link.  Of course, $\hat{W}_{\mathrm{el}}$ can be calculated explicitly, though the integral may be awkward to do analytically, and it depends on the representation of the gauge field on qubits.

%For the example of quantum link models \cite{CW97}, the gauge-field operators no longer satisfy $[\hat{U}^{ab}_{\mathbf{x},\mathbf{e}},\hat{U}^{cd}_{\mathbf{x},\mathbf{e}}] = 0$ on the same link.  This was satisfied for the original gauge theory, but we also had infinite-dimensional Hilbert spaces for each link.  In quantum link models, the Hilbert space at each link is finite, and $\hat{U}^{ab}_{\mathbf{x},\mathbf{e}}$ are operators on a finite-dimensional Hilbert space.  For example, for a quantum link model for $\mathrm{SU}(2)$, the 

\section{Discussion}
We saw that taking $\delta t = a$ is sufficient for quantum simulations of QFT, at least for bosonic QFTs.  This is important because, in some experiments simulating QFTs, the ratio $\delta t/a$ actually cannot be too small for technical reasons.  Another benefit is a new quantum information perspective on the discrete-time path integral approach to lattice QFT, which is the approach of choice in \emph{classical} simulations of QFT.  We saw that the path integral expression  is equivalent to a quantum circuit, so in contrast to the usual approach in quantum simulation of QFT, it is sometimes natural to use the discrete-time Lagrangian formulation of lattice QFT.  In fact, a priori there is no reason why the continuous-time Hamiltonian formulation is physically more fundamental than the discrete-time Lagrangian formulation, which puts time and space on an even footing.  Of course, the connection between path integrals in Euclidean Lagrangian lattice QFT and the imaginary-time transfer matrix is well known.  The crucial point here was that for scalar and non-abelian gauge fields the real-time transfer matrix, and hence the real-time path integral, are exactly equal to finite-depth local quantum circuits.  A similar idea does not work so easily for fermions because the transfer matrix does not factor exactly into a finite-depth local circuit for, say, Wilson fermions.  

It is also possible to go beyond circuits that arise from lattice QFT, as we saw with the Shift circuit.  This suggests that there are more possibilities for simulating QFT via quantum computers than arise from lattice QFT.

We focused on dynamics, but there are other key aspects of simulation, such as initial state preparation and measurement.  Since the circuits here approximate the dynamics of QFTs, it should be sufficient to use any approximation of the ground or initial state of the QFT for simulations, as long as both the circuit dynamics and the initial state have the right continuum limits.  For scattering processes in $\phi^4$ theory, state preparation and measurement were dealt with in \cite{JLP12}.  There are other details that cannot be swept under the rug, such as rates of convergence, as they affect the computational complexity of simulations.  One way to deal with this is to use effective field theory to estimate the error arising from the discretization, as in \cite{JLP12}, or to rely on results from Euclidean lattice QFT quantifying discretization errors and improving the action to reduce errors.

As a final point, in practice it would be interesting to do renormalization via gradient descent, especially in cases where perturbative methods no longer work.  This could work as follows.  Physical parameters at a fixed energy scale are defined by certain processes, e.g., $\lambda_{\mathrm{phys}}$ can be defined by an amplitude for two particles scattering to two particles, giving $-i\lambda_{\mathrm{phys}}$ \cite{PeskinSchroeder}.  Now, if we know the physical parameters $g^i_{\mathrm{phys}}$ at some energy scale (maybe from experiments).  Then to find the right bare parameters $g_0^i$ for our simulator at fixed lattice spacing, we simulate the processes that define the physical parameters getting $g^i_{\mathrm{sim}}(g_0^j)$, which depend on the bare parameters $g_0^i$.  Then we vary the bare parameters to minimize 
\begin{equation}
C=\sum_i[g^i_{\mathrm{phys}}-g^i_{\mathrm{sim}}(g^j_0)]^2.  
\end{equation}
To do this, we can approximate $\partial C/\partial g^i_0$ by running the simulations with slightly shifted parameters $g^i_0+\delta g_0^i$.  Following the usual recipe for gradient descent, we update our bare parameters:
\begin{equation}
g^i_0\rightarrow g^i_0-\eta\partial C/\partial g^i_0, 
\end{equation}
where $\eta>0$ is a free parameter we choose empirically.  Of course, this comes with all the fineprint associated with gradient descent, e.g., local minima.  Nevertheless, even if the procedure is computationally expensive (we have no obvious guide to the computational complexity), it is a one-time cost:\ the best values of the bare parameters for each lattice spacing need only be found once.

\section*{Note added}
Close to completion of this work, a preprint \cite{19BMP} was posted on the arXiv that also looks at scattering in discrete-time, a topic discussed in appendix \ref{app:int-fields}.  In contrast to our case, in \cite{19BMP} a discretization of the Thirring model was studied.

\acknowledgments
TCF was supported by the Australian Research Council Centres of Excellence for Engineered Quantum Systems (EQUS, CE170100009).  The authors would like to thank Tom Stace, Tobias Osborne and Dmytro Bondarenko for helpful discussions and Sinduja Suresh for help with figures.

\bibliographystyle{unsrt}
%\bibliography{References2}

\appendix
\section{Relation of the Strang-split circuit to the path integral}
\label{sec:Strang-proof}
In this and the following sections, it is sometimes more convenient to use the dimensionless versions of the on-site scalar field operators $X_{\mathbf{n}}$ and $P_{\mathbf{n}}$, which satisfy
\begin{equation}
\left[X_{\mathbf{m}},P_{\mathbf{n}}\right]=i\delta_{\mathbf{m},\mathbf{n}}
\end{equation}
and $[X_{\mathbf{m}},X_{\mathbf{n}}]=[P_{\mathbf{m}},P_{\mathbf{n}}]=0$.
In terms of these, the discrete quantum field operators are
\begin{equation}
 \begin{split}
  X_{\mathbf{n}} & = a^{(d-1)/2}\phi_{\mathbf{n}}\\
   P_{\mathbf{n}} & = a^{(d+1)/2}\pi_{\mathbf{n}}.
 \end{split}
\end{equation}

In this section, we want to relate the Strang-split circuit to the discrete-time path integral.  Let us start by introducing a basis describing the field configurations on the lattice at time $\eta$ such that
\begin{equation}
 X_{\mathbf{n}}|x(\eta)\rangle = x_{\mathbf{n}}(\eta)|x(\eta)\rangle,
\end{equation}
so here $x_{\mathbf{n}}(\eta)$ is a scalar, so we are using upper case letters for operators and lower case letters for scalars.  Since $X_{\mathbf{n}}$ is proportional to $\phi_{\mathbf{n}}$, these states are also eigenstates of the field operators $\phi_{\mathbf{n}}$.  Furthermore, we have
\begin{equation}
\begin{split}
& |x(\eta)\rangle = \prod_{\mathbf{n}}|x_{\mathbf{n}}(\eta)\rangle\\
\openone & =\int\mathrm{d}x(\eta)|x(\eta)\rangle\langle x(\eta)|,
\end{split}
\end{equation}
where $\int\mathrm{d}x(\eta)=\int\prod_{\mathbf{n}}\mathrm{d}x_{\mathbf{n}}(\eta)$.  Note that if the spatial lattice is finite, then everything here is mathematically well defined.  If the lattice is however infinite, i.e., $\mathbb{Z}^d$, then these expressions are only formal.

We have the initial and final field configurations $|x(0)\rangle$ and $|x(\tau)\rangle$ at time $0$ and time $\tau$ respectively.  The amplitude for one to evolve into the other is
\begin{equation}
 \langle x(\tau)|U^{\tau}|x(0)\rangle = \langle x(\tau)|(W_XW_PW_X)^{\tau}|x(0)\rangle,
\end{equation}
where we have
\begin{equation}
\begin{split}
 W_P & = \prod_{\mathbf{n}\in \mathbb{Z}^d} \exp\!\left[\frac{-i}{2}
 P_{\mathbf{n}}^2\right]\\
 W_X & = \prod_{\mathbf{n}\in \mathbb{Z}}\exp\!\left[\frac{-i}{2} V(X_{\mathbf{n}})\right],
 \end{split}
\end{equation}
where
\begin{equation}
 V(X_{\mathbf{n}})  = (d+m^2a^2/2)X_{\mathbf{n}}^2-\sum_{\mathbf{e}\in\mathcal{N}}X_{\mathbf{n}+\mathbf{e}}X_{\mathbf{n}}+\frac{a^{1-d}\lambda}{4!}X_{\mathbf{n}}^4.
\end{equation}
Inserting factors of $\openone=\int\mathrm{d}x(\eta)|x(\eta)\rangle\langle x(\eta)|$, we get
\begin{equation}
 \langle x(\tau)|U^{\tau}|x(0)\rangle = \int\prod_{\eta=1}^{\tau-1}\mathrm{d}x(\eta)\prod_{\nu=0}^{\tau-1}\langle x(\nu+1)|W_XW_PW_X|x(\nu)\rangle.
\end{equation}
We can simplify this via
\begin{equation}
\begin{split}
& \langle x(\nu+1)|W_XW_PW_X|x(\nu)\rangle\\
& = \langle x(\nu+1)|W_P|x(\nu)\rangle e^{-\frac{i}{2}\sum_{\mathbf{n}}[V[x_{\mathbf{n}}(\nu+1)]+V[x_{\mathbf{n}}(\nu)]]},
\end{split}
\end{equation}
where now $V[x_{\mathbf{n}}(\nu)]$ is a function of the scalars $x_{\mathbf{n}}(\nu)$.  Also, we have
\begin{equation}
\begin{split}
\langle x(\nu+1)|W_P|x(\nu)\rangle & = \prod_{\mathbf{n}}\langle x_{\mathbf{n}}(\nu+1)| e^{\frac{-i}{2}P_{\mathbf{n}}^2}|x_{\mathbf{n}}(\nu) \rangle
 \end{split}
\end{equation}
Since $P_{\mathbf{n}}$ is the canonically conjugate variable to $X_{\mathbf{n}}$, we can insert the identity in the form of an integral over eigenvectors of $P_{\mathbf{n}}$, given by (dropping the ${\mathbf{n}}$ indices for now)
\begin{equation}
 \openone = \int\frac{\mathrm{d}p}{2\pi}|p\rangle\langle p |,
\end{equation}
where the integral is over $\mathbb{R}$ and
\begin{equation}
 |p\rangle = \int\frac{\mathrm{d}p}{2\pi}e^{ip z}|z\rangle.
\end{equation}
Then we get
\begin{equation}
\begin{split}
\langle y| e^{\frac{-i}{2}P^2}|z \rangle & = \int\frac{\mathrm{d}p}{2\pi}\langle y|p\rangle\langle p|z\rangle e^{\frac{-i}{2}p^2}\\
 & = \int\frac{\mathrm{d}p}{2\pi} \exp\left(\frac{-i}{2} p^2+ip(y-z)\right)\\
  & = \sqrt{\frac{i}{2\pi}}\exp\left(\frac{i}{2}\left[y-z\right]^2\right)
 .
 \end{split}
\end{equation}
Putting this together, we get that
\begin{equation}
 \langle x(\tau)|U^{\tau}|x(0)\rangle = \int D[x]\exp\left(i\sum_{\eta=0}^{\tau-1}\sum_{\mathbf{n}}L(\eta,\mathbf{n})\right)
\end{equation}
where
\begin{equation}
\begin{split}
L(\eta,\mathbf{n}) & = \frac{1}{2}\left[x_{\mathbf{n}}(\eta+1)-x_{\mathbf{n}}(\eta)\right]^2\\
& -\frac{1}{2}[V[x_{\mathbf{n}}(\eta+1)]+V[x_{\mathbf{n}}(\eta)]]
\end{split}
\end{equation}
and
\begin{equation}
D[x]=\prod_{\nu=1}^{\tau-1}\left(\prod_{\mathbf{n}}\mathrm{d}x_{\mathbf{n}}(\nu)\right)\left(\prod_{\mu=1}^{\tau-1}\prod_{\mathbf{m}}\sqrt{\frac{i}{2\pi}}\right).
\end{equation}
Subbing in $x_{\mathbf{n}}(\nu)=a^{(d-1)/2}\varphi_{\nu,\mathbf{n}}$, we finally we see that
\begin{equation}
 \langle \varphi_{\mathrm{f}}|U^{\tau}|\varphi_{\mathrm{i}}\rangle = \int \mathcal{D}(\varphi)e^{iS(\varphi)},
\end{equation}
where the action $S(\varphi)$ is given by
\begin{equation}
S(\varphi) = \sum^{\tau-1}_{\nu=0}\delta t\sum_{\mathbf{n}}a^d\mathcal{L}(\nu,\mathbf{n}),
\end{equation}
with 
\begin{equation}
\begin{split}
\mathcal{L}(\nu,\mathbf{n}) & = \frac{(\varphi_{\tau+1,\mathbf{n}}-\varphi_{\tau,\mathbf{n}})^2}{2\delta t^2}- \frac{\mathcal{V}(\varphi_{\nu,\mathbf{n}})+\mathcal{V}(\varphi_{\nu,\mathbf{n}})}{2}
\end{split}
\end{equation}
and
\begin{equation}
\begin{split}
\mathcal{V}(\varphi_{\nu,\mathbf{n}})= \frac{1}{2}(\nabla_a \varphi_{\nu,\mathbf{n}})^2+\frac{m^2}{2}\varphi_{\nu,\mathbf{n}}^2+\frac{\lambda}{4!}\varphi_{\nu,\mathbf{n}}^4.
\end{split}
\end{equation}
Finally, the measure is given by
\begin{equation}
 \mathcal{D}(\varphi)=\prod_{\eta=1}^{\tau-1}\prod_{\mathbf{n}}\left(\sqrt{\frac{ia^{d-1}}{2\pi}}\mathrm{d}\varphi_{\eta,\mathbf{n}}\right).
\end{equation}

\section{The Shift circuit}
Recall that the Shift circuit has evolution operator given by $U=W_XW_PW_X$, but with
\begin{equation}
\begin{split}
 W_X & = \prod_{\mathbf{n}\in \mathbb{Z}^d} \exp\!\left[ia^{d-1}\!\!\left(\frac{M}{2^{D}}\sum_{\mathbf{e}\in N}\phi_{\mathbf{n}}\phi_{\mathbf{n}+\mathbf{e}}+\frac{\lambda a^2}{4!}\phi_{\mathbf{n}}^4\!\right)\!\right]\\
 W_P & = \prod_{\mathbf{n}\in \mathbb{Z}^d} \exp\!\left[-\frac{i\pi a^{d-1}}{4}(\phi_{\mathbf{n}}^2+a^2\pi_{\mathbf{n}}^2)\right],
 \end{split}
\end{equation}
where $D=d+1$, $M=1-m^2a^2/2$ ($m$ is the mass), and $N$ are lattice vectors $\mathbf{e}$ with components $e_i=\pm 1$.  It is sometimes convenient to use the dimensionless fields and to write this as
\begin{equation}
\begin{split}
 W_X & = \prod_{\mathbf{n}\in \mathbb{Z}^d} \exp\!\left[\frac{iM}{2^{D}}\sum_{\mathbf{e}\in N}X_{\mathbf{n}}X_{\mathbf{n}+\mathbf{e}}+\frac{a^{d-1}\lambda}{4!}X_{\mathbf{n}}^4\right]\\
 W_P & = \prod_{\mathbf{n}\in \mathbb{Z}^d} \exp\!\left[-\frac{i\pi}{4}(X_{\mathbf{n}}^2+P_{\mathbf{n}}^2)\right].
 \end{split}
\end{equation}

\subsection{Continuum limit in the free case}
In the free case with $\lambda=0$, it is straightforward to take the continuum limit.  The circuit evolving the system over one timestep is 
\begin{equation}\label{eq:free_unitary}
 U_0=W^0_XW_PW^0_X,
\end{equation}
where again each individual $W^0_X$ or $W_P$ is a product of commuting local unitaries:
\begin{equation}
\begin{split}
 W^0_X & = \prod_{\mathbf{n}\in \mathbb{Z}^d} \exp\!\left[\frac{iM}{2^{D}}\sum_{\mathbf{e}\in N}X_{\mathbf{n}}X_{\mathbf{n}+\mathbf{e}}\right]\\
 W_P & = \prod_{\mathbf{n}\in \mathbb{Z}^d} \exp\!\left[-\frac{i\pi}{4}(X_{\mathbf{n}}^2+P_{\mathbf{n}}^2)\right].
 \end{split}
\end{equation}
Let us see how these unitaries act.  First, $W^0_X$ leaves $X_\mathbf{n}$ invariant.  So the non trivial effects of the operators are
\begin{equation}\label{eq:usef1}
   W_X^{0\dagger}P_\mathbf{n} W^0_X = P_\mathbf{n} +\frac{M}{2^d}\sum_{\mathbf{e}\in N}(X_{\mathbf{n}+\mathbf{e}}+X_{\mathbf{n}-\mathbf{e}})
   \end{equation}
   and
   \begin{equation}\label{eq:usef2}
 \begin{split}
W_P^{\dagger}X_\mathbf{n} W_P & =  P_\mathbf{n}\\
W_P^{\dagger}P_\mathbf{n} W_P & = -X_\mathbf{n}.
 \end{split}
\end{equation}

To take the continuum limit and for solving the circuits, it will be useful to switch to momentum space.  For any operators $A_{\mathbf{n}}$, we define their momentum space representation via
\begin{equation}
 A(\mathbf{p})=a^d\sum_{\mathbf{n}\in\mathbb{Z}^d}e^{-i\mathbf{p}.\mathbf{n}a}A_{\mathbf{n}},
\end{equation}
where $\mathbf{p}$ denotes a momentum vector, with components satisfying $p_i\in(-\pi/a,\pi/a]$.  Next we need the identity
\begin{equation}
a^d\sum_{\mathbf{n}\in\mathbb{Z}^d}e^{-i\mathbf{p}.\mathbf{n}a}A_{\mathbf{n}+\mathbf{m}}= e^{i\mathbf{p}.\mathbf{m}a}A(\mathbf{p}).
\end{equation}
Then we get that
\begin{equation}\label{eq:1}
 \begin{split}
  W_X^{0\dagger}P(\mathbf{p}) W^0_X & = P(\mathbf{p}) + c(\mathbf{p})X(\mathbf{p})
 \end{split}
\end{equation}
and
\begin{equation}\label{eq:2}
 \begin{split}
 W_P^{\dagger}X(\mathbf{p}) W_P & = P(\mathbf{p})\\
W_P^{\dagger}P(\mathbf{p}) W_P & = -X(\mathbf{p}).
 \end{split}
\end{equation}
where we have introduced
\begin{equation}
 c(\mathbf{p})=M\prod_{i=1}^d\cos(p_ia).
\end{equation}
It will be helpful to note how $c(\mathbf{p})$ behaves.  For small momenta, with each $p_i\ll 1/a$, we can Taylor expand to see that $c(\mathbf{p}) = 1 -(\mathbf{p}^2+m^2)a^2/2+ O(a^4)$.

Using equation (\ref{eq:free_unitary}) and equations (\ref{eq:1}) and (\ref{eq:2}), we find that the evolution in momentum space is
\begin{equation}
\begin{split}
U_0^{\dagger}X(\mathbf{p})U_0 & = c(\mathbf{p})X(\mathbf{p})+ P(\mathbf{p})\\
U_0^{\dagger}P(\mathbf{p})U_0 & = c(\mathbf{p})P(\mathbf{p})+(c(\mathbf{p})^2-1)X(\mathbf{p}).
\end{split}
\end{equation}
In terms of the field operators, for small momentum, we see that
\begin{equation}
\begin{split}
U_0^{\dagger}\phi(\mathbf{p})U_0 & = \phi(\mathbf{p})+\pi(\mathbf{p})a + O(a^2)\\
U_0^{\dagger}\pi(\mathbf{p})U_0 & = \pi(\mathbf{p})-a(\mathbf{p}^2+m^2)\phi(\mathbf{p}) + O(a^2).
\end{split}
\end{equation}
This allows us to find time derivatives in the continuum limit.  In the Heisenberg picture we have, e.g., $\phi(t+a,\mathbf{p})=U_0^{\dagger}\phi(t,\mathbf{p})U_0$, where we used that $\delta t = a$.  Then we can find the time derivative of the field as the lattice spacing $a\propto \delta t$ goes to zero:
\begin{equation}
 \partial_t \phi(t,\mathbf{p}) = \lim_{a\rightarrow 0}\frac{U_0^{\dagger}\phi(t,\mathbf{p})U_0-\phi(t,\mathbf{p})}{a}.
\end{equation}
This gives us back exactly the equations of motion for the Klein-Gordon field in momentum space:
\begin{equation}
\begin{split}
 \partial_t \phi(t,\mathbf{p}) & = \pi(t,\mathbf{p})\\
\partial_t \pi(t,\mathbf{p}) & = -(\mathbf{p}^2+m^2)\phi(t,\mathbf{p}).
\end{split}
\end{equation}
So we see that for low momenta, or equivalently physics on length scales that are large compared to the lattice spacing, we have the usual scalar field dynamics.

\subsection{Motivation}
\label{sec:Shift-motivation}
The reason for our choice of circuit is that it is a natural discretization of the continuum model when $m=\lambda=0$ in $d=1$.  This is because, in that setting, the continuum model can be solved in terms of left- and right-moving fields, which also applies to this circuit.  To see how this works, let us make a brief digression to continuous spacetime.  There we have fields $\phi(\mathbf{x})$ and $\pi(\mathbf{x})$ ($\mathbf{x}\in\mathbb{R}$ is the spatial coordinate) obeying
\begin{equation}
\begin{split}
 [\phi(\mathbf{x}),\pi(\mathbf{y})] & =i\delta (\mathbf{x}-\mathbf{y})\\
 [\phi(\mathbf{x}),\phi(\mathbf{y})] & =[\pi(\mathbf{x}),\pi(\mathbf{y})]  =0.
 \end{split}
\end{equation}
The dynamics is given by the Hamiltonian
\begin{equation}
H = \frac{1}{2}\int\!\!\mathrm{d}\mathbf{x}\left[ \pi^2(\mathbf{x})+\left(\frac{\partial\phi(\mathbf{x})}{\partial \mathbf{x}}\right)^2\right].
\end{equation}
To solve this, we introduce left and right-moving fields, given by
\begin{equation}
 \begin{split}
\pi_{L}(\mathbf{x}) & = \frac{1}{2}\left(\pi(\mathbf{x})+\partial_{\mathbf{x}}\phi(\mathbf{x})\right)\\
\pi_{R}(\mathbf{y}) & = \frac{1}{2}\left(\pi(\mathbf{x})-\partial_{\mathbf{x}}\phi(\mathbf{x})\right),
\end{split}
\end{equation}
which satisfy the commutation relations
\begin{equation}
\begin{split}
 [\pi_{L}(\mathbf{x}),\pi_{L}(\mathbf{y})] & =\frac{i}{2}\partial_{\mathbf{x}}\delta (\mathbf{x}-\mathbf{y})\\
 [\pi_{R}(\mathbf{x}),\pi_{R}(\mathbf{y})] & =-\frac{i}{2}\partial_{\mathbf{x}}\delta (\mathbf{x}-\mathbf{y})\\
 [\pi_{R}(\mathbf{x}),\pi_{L}(\mathbf{y})] & =0.
 \end{split}
\end{equation}
By using $\dot{A}(t)=i[H,A(t)]$, we see that the dynamics in the Heisenberg picture for the free field is simply
\begin{equation}\label{eq:L-R-cont}
 \begin{split}
\pi_{L}(t,\mathbf{x}) & = \pi_{L}(0,\mathbf{x}+t)\\
\pi_{R}(t,\mathbf{x}) & = \pi_{R}(0,\mathbf{x}-t).
\end{split}
\end{equation}
In other words, as the name suggests, left-moving fields move left and right-moving fields move right.

Returning to discrete spacetime, the Shift quantum circuit is a simple discretization of this when restricted to $d=1$ and $m=0$ with $\lambda=0$.  To see this, we can define lattice left-moving and right-moving fields $\pi_{L,n}$ and $\pi_{R,n}$ by
\begin{equation}
 \begin{split}
\pi_{L,n} & = \frac{1}{2}\left(\pi_{n}+\frac{\phi_{n+1}-\phi_{n-1}}{2a}\right)\\
\pi_{R,n} & = \frac{1}{2}\left(\pi_{n}-\frac{\phi_{n+1}-\phi_{n-1}}{2a}\right),
\end{split}
\end{equation}
which obey the commutation relations
\begin{equation}
 \begin{split}
\left[\pi_{L,m},\pi_{R,n}\right] & =0\\
\left[\pi_{L,m},\pi_{L,n}\right] & =-\frac{i}{4a^{2}}\left(\delta_{m,n+1}-\delta_{m,n-1}\right)\\
\left[\pi_{R,m},\pi_{R,n}\right] & =\frac{i}{4a^{2}}\left(\delta_{m,n+1}-\delta_{m,n-1}\right).
\end{split}
\end{equation}
The operators $\pi_{L,n}$ and $\pi_{R,n}$ tend in the continuum limit to the continuum right and left-movers $\pi_{R}(\mathbf{x})$ and $\pi_{L}(\mathbf{x})$ with the correct commutation relations.  Furthermore, with $m=0$ (and hence $M=1$) the unitary $U_0$ in $d=1$ (equation (\ref{eq:free_unitary})) shifts $\pi_{L,n}$ left and $\pi_{R,n}$ right over each timestep, i.e.,
\begin{equation}
\begin{split}
 U^{\dagger}_0\pi_{L,m}U^{\dagger}_0 & = \pi_{L,m+1}\\
 U^{\dagger}_0\pi_{R,m}U^{\dagger}_0 & = \pi_{R,m-1},
 \end{split}
\end{equation}
which is a simple discretization of equation (\ref{eq:L-R-cont}) and follows from equations (\ref{eq:usef1}) and (\ref{eq:usef2}).

\subsection{Solution in the free case}\label{sec:mainsol}
We can solve the free Shift circuit $U_0$ defined in equation (\ref{eq:free_unitary}).  We could also solve the Strang-split circuit, but the solution is already known from lattice QFT albeit in the Wick-rotated imaginary time setting, but the basic idea is similar.

In momentum space, the evolution of the field over one timestep is given by
\begin{equation}\label{eq:dyno}
\begin{split}
U_0^{\dagger}\phi(\mathbf{p})U_0 & = c(\mathbf{p})\phi(\mathbf{p})+a\pi(\mathbf{p})\\
U_0^{\dagger}\pi(\mathbf{p})U_0 & = c(\mathbf{p})\pi(\mathbf{p})+\frac{[c(\mathbf{p})^2-1]}{a}\phi(\mathbf{p}),
\end{split}
\end{equation}
where
\begin{equation}
c(\mathbf{p})= M\prod_{i=1}^d\cos(p_ia).
\end{equation}

To solve equation (\ref{eq:dyno}), we need to find the annihilation operator
\begin{equation}\label{eq:free_part}
 b_{\mathbf{p}}=\alpha(\mathbf{p})\phi(\mathbf{p})+\beta(\mathbf{p})\pi(\mathbf{p})
\end{equation}
satisfying
\begin{equation}\label{eq:solve_c}
 U_0^{\dagger}b_{\mathbf{p}}U_0 = e^{-i\theta(\mathbf{p})a}b_\mathbf{p},
\end{equation}
while also obeying
\begin{equation}
[b_{\mathbf{p}},b_{\mathbf{q}}^{\dagger}]=(2\pi)^d\delta^{(d)}(\mathbf{p}-\mathbf{q}).
\end{equation}
So our goal is to find $\alpha(\mathbf{p})$, $\beta(\mathbf{p})$ and $\theta(\mathbf{p})$.
The most straightforward way to do this is to write equation (\ref{eq:solve_c}) as a matrix equation.  To do this, we note that the dynamics of the field in equation (\ref{eq:dyno}) can be specified by the matrix
\begin{equation}
 \begin{pmatrix}
  c(\mathbf{p}) & \frac{c(\mathbf{p})^2-1}{a}\\
  a & c(\mathbf{p})
 \end{pmatrix},
\end{equation}
where the vector $(1,0)^{\mathrm{T}}$ represents $\phi(\mathbf{p})$ and $(0,1)^{\mathrm{T}}$ represents $\pi(\mathbf{p})$.
This is symmetric under $\mathbf{p}\rightarrow - \mathbf{p}$.
In matrix form, equation (\ref{eq:solve_c}) becomes the eigenvalue equation
\begin{equation}\label{eq:lmnrt}
\begin{pmatrix}
  c(\mathbf{p}) & \frac{c(\mathbf{p})^2-1}{a}\\
  a & c(\mathbf{p})
 \end{pmatrix}
  \begin{pmatrix}
  \alpha(\mathbf{p}) \\
  \beta(\mathbf{p})
 \end{pmatrix}
 = e^{-i\theta(\mathbf{p}) a}\begin{pmatrix}
  \alpha(\mathbf{p}) \\
  \beta(\mathbf{p})
 \end{pmatrix}.
\end{equation}
(The other eigenvalue $e^{i\theta(\mathbf{p}) a}$ turns out to correspond to the creation operator $b_{\mathbf{p}}^{\dagger}$.)
In order to get real solutions for $\theta(\mathbf{p})$ (which is the circuit's analogue of energy), we need that $-1<c(\mathbf{p})<1$, which is true if $m>0$ since then $M<1$.

Solving equation (\ref{eq:lmnrt}) gives us
\begin{equation}
 e^{-i\theta(\mathbf{p}) a} = c(\mathbf{p}) \pm\sqrt{c(\mathbf{p})^2-1}.
\end{equation}
Since we know that $0< c(\mathbf{p})< 1$, we can rewrite this as
\begin{equation}
 e^{-i\theta(\mathbf{p}) a} = c(\mathbf{p}) - i\sqrt{1-c(\mathbf{p})^2},
\end{equation}
where the choice of the minus sign corresponds to positive $\theta(\mathbf{p})$.  The other eigenvalue $e^{i\theta(\mathbf{p}) a}$, which has a plus sign, corresponds to the eigenvector $(\alpha^*(\mathbf{p}), \beta^*(\mathbf{p}))^T$, which describes the creation operator $b_{-\mathbf{p}}^{\dagger}$.

The eigenvalue equation does not fix $\alpha(\mathbf{p})$ and $\beta(\mathbf{p})$ uniquely.  But we also have the constraint that $[b_{\mathbf{p}},b_{\mathbf{q}}^{\dagger}]=(2\pi)^d\delta^{(d)}(\mathbf{p}-\mathbf{q})$, which implies that
\begin{equation}\label{eq:useful}
 \alpha(\mathbf{p})\beta^*(\mathbf{p})-\alpha^*(\mathbf{p})\beta(\mathbf{p})=-i.
\end{equation}
Then we can satisfy this by choosing
\begin{equation}
 \begin{split}
  \alpha(\mathbf{p}) & = \sqrt{\frac{\sin[\theta(\mathbf{p})a]}{2a}}\\
  \beta(\mathbf{p}) & = i\sqrt{\frac{a}{2\sin[\theta(\mathbf{p})a]}},
   \end{split}
\end{equation}
which also solve equation (\ref{eq:lmnrt}).

Next we can invert for $b_{\mathbf{p}}$ to find the field operators.  We have
\begin{equation}
\begin{split}
 b_{\mathbf{p}}^{\dagger} & =\alpha^*(\mathbf{p})\phi^{\dagger}(\mathbf{p})+\beta^*(\mathbf{p})\pi^{\dagger}(\mathbf{p})\\
 & =\alpha^*(\mathbf{p})\phi(-\mathbf{p})+\beta^*(\mathbf{p})\pi(-\mathbf{p}),
 \end{split}
\end{equation}
where we used that, e.g., $\phi^{\dagger}(\mathbf{p})=\phi(-\mathbf{p})$.  Also, noting that both $\alpha(\mathbf{p})$ and $\beta(\mathbf{p})$ are symmetric under $\mathbf{p}\rightarrow-\mathbf{p}$, which follows because $\theta(\mathbf{p})$ is also symmetric, we get
\begin{equation}
\begin{split}
 b_{-\mathbf{p}}^{\dagger} & =\alpha^*(\mathbf{p})\phi(\mathbf{p})+\beta^*(\mathbf{p})\pi(\mathbf{p})
 \end{split}
\end{equation}
The trick now is to use this equation as well as equation (\ref{eq:free_part}) to isolate $\phi(\mathbf{p})$.  So we get
\begin{equation}
\begin{split}
 \beta^*(\mathbf{p})b_{\mathbf{p}} - \beta(\mathbf{p})b_{-\mathbf{p}}^{\dagger} & =\left[\beta^*(\mathbf{p})\alpha(\mathbf{p})-\beta(\mathbf{p})\alpha^*(\mathbf{p})\right]\phi(\mathbf{p})\\
 & =-i\phi(\mathbf{p}),
 \end{split}
\end{equation}
where the last line followed from equation (\ref{eq:useful}).  Substituting in $\beta(\mathbf{p})=i(2\omega(\mathbf{p}))^{-1/2}$, we get
\begin{equation}
\begin{split}
\phi(\mathbf{p})=\frac{1}{\sqrt{2\omega(\mathbf{p})}}\left[b_{\mathbf{p}} + b_{-\mathbf{p}}^{\dagger}\right].
 \end{split}
\end{equation}
Fourier transforming gives 
\begin{equation}
\begin{split}
\phi(\mathbf{x}) & =\int\!\frac{d^dp}{(2\pi)^d}\frac{e^{i\mathbf{p}.\mathbf{x}}}{\sqrt{2\omega(\mathbf{p})}}\left[b_{\mathbf{p}} + b_{-\mathbf{p}}^{\dagger}\right]\\
& =\int\!\frac{d^dp}{(2\pi)^d}\frac{1}{\sqrt{2\omega(\mathbf{p})}}\left[e^{i\mathbf{p}.\mathbf{x}}b_{\mathbf{p}} + e^{-i\mathbf{p}.\mathbf{x}}b_{\mathbf{p}}^{\dagger}\right].
 \end{split}
\end{equation}
Finally, going to the Heisenberg picture, and using that
\begin{equation}
\begin{split}
 U_0^{\dagger}b_{\mathbf{p}}U_0 & = e^{-i\theta(\mathbf{p})a}b_\mathbf{p}\\
 U_0^{\dagger}b_{\mathbf{p}}^{\dagger}U_0 & = e^{i\theta(\mathbf{p})a}b_\mathbf{p}^{\dagger},
 \end{split}
\end{equation}
gives
\begin{equation}\label{eq:phi_again}
\begin{split}
\phi(x) & = \int\!\frac{d^dp}{(2\pi)^d}\frac{1}{\sqrt{2\omega(\mathbf{p})}}\left(e^{-ip_s.x}b_{\mathbf{p}}+e^{ip_s.x}b_{\mathbf{p}}^{\dagger}\right),
\end{split}
\end{equation}
where $p_s=(\theta(\mathbf{p}),\mathbf{p})$.
Similarly, we find an expression for $\pi(x)$, which is
\begin{equation}
\pi(x)  = -i\int\!\frac{d^dp}{(2\pi)^d}\sqrt{\frac{\omega(\mathbf{p})}{2}}\left(e^{-ip_s.x}b_{\mathbf{p}}-e^{ip_s.x}b_{\mathbf{p}}^{\dagger}\right).
\end{equation}

\begin{figure}[!ht]
{\centering
\resizebox{7.5cm}{!}{\includegraphics{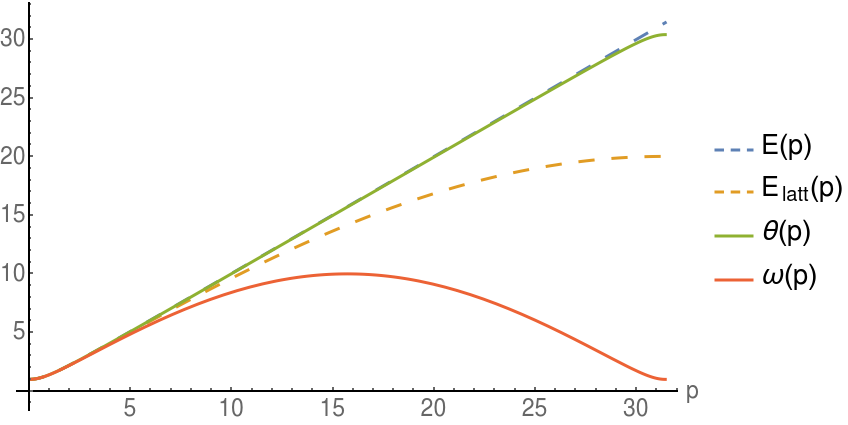}} \caption[Contour]{Comparison of energies of the Shift circuit in $d=1$.  $E(p)=\sqrt{p^2+m^2}$ is the continuum dispersion relation, and $E_{\mathrm{latt}}(p)=\sqrt{m^2 + 4\sin^2[pa/2]/a^2}$ is the dispersion relation for the lattice Hamiltonian with $\lambda=0$ from equation (\ref{eq:lattHam}).  For the Shift circuit, there are two functions that play the role of energy, $\theta(p)$ appears in the phase picked up by annihilation operators, and $\omega(p)$ appears in equal-time correlation functions.  These are plotted with lattice spacing taken to be $0.1$ and $m=1$.  The plot is over half the momentum range $[0,\pi/a]$ since it is symmetric.  For this model $\delta t= a$.  In this case, $\omega(p)$ is a poor approximation to $E(p)$ for large $p$, but, on the other hand, $\theta(p)$ is staggeringly close to $E(p)$. \label{fig:energies_alt}}
}
\end{figure}

The free circuit vacuum is given by $|0\rangle$, the state annihilated by all $b_{\mathbf{p}}$.  Furthermore, vacuum correlations are described by the equal-time propagator:
\begin{equation}
\langle 0|\phi(x)\phi(y)|0\rangle = \!\int\! \frac{\mathrm{d}^dp}{(2\pi)^d}\frac{e^{i\mathbf{p}.(\mathbf{x}-\mathbf{y})}}{2\omega(\mathbf{p})},
\end{equation}
where $x^0=y^0$.  As $\omega(\mathbf{p})\rightarrow E(\mathbf{p})$ for small $\mathbf{p}a$, the circuit equal-time propagator reproduces the continuum version for small momenta.  This means that the circuit vacuum approximates the continuum vacuum well at length scales that are large compared to $a$.

In Hamiltonian lattice QFT or continuum QFT, the energy that appears in the exponents in the field operators and in the factor after the integral in the field operator are the same.  In contrast, here (and in the discrete-time Lagrangian formalism of lattice QFT \cite{Smit02}) we have the discrete energy $\theta(\mathbf{p})$ in the exponent and a different factor $\omega(\mathbf{p})=\sin[\theta(\mathbf{p})a]/a$ in the denominator in the integral.  Nevertheless, for small lattice momenta, both coincide, and the expressions look like the familiar continuum field operators.  These two functions, $\theta(\mathbf{p})$ and $\omega(\mathbf{p})$, are plotted in figure \ref{fig:energies_alt}.  Interestingly, $\theta(\mathbf{p})$ is an extremely good approximation to $E(\mathbf{p})$, whereas $\omega(\mathbf{p})$ becomes a poor approximation for large $p$.

However, there is a slight problem with this model.  The equal-time propagator depends on $\omega(\mathbf{p})=\sin[\theta(\mathbf{p})a]=\sqrt{1-M^2\prod_i\cos^2(p_ia)}$ in the denominator, which has small values, not only for $|\mathbf{p}a|\ll 1$, but also for high momenta $|p_ia|\sim \pi$, which is analogous to fermion doubling.  To mitigate this problem, we will modify the interaction from the point interaction $\phi_{\mathbf{n}}^4$ to something averaged over a few sites thus averaging out high momentum contributions to scattering amplitudes.  We look at this problem and its solution in more detail in section \ref{app:One_loop}.  Another option would be to modify the free circuit along the lines of Wilson's solution to the fermion doubling problem, i.e., by introducing an artificial mass term that only affects high momentum particles.

\section{Feynman propagator}\label{app:Feyn_prop}
In this section, we allow a general $\delta t$, which may or may not equal $a$, since the argument applies either way.  Our goal is to find the discrete spacetime Feynman propagator:
\begin{equation}\label{eq:asdf2}
 D_F(x-y) =\langle 0|\mathcal{T}\!\left[\phi(x)\phi(y)\right]\!|0\rangle
 \end{equation}
 where $\mathcal{T}$ is the time-ordering operator.  We will show that the Feynman propagator can be written as
\begin{equation}\label{eq:asdf}
\begin{split}
 D_F(x-y) & = \frac{\delta t^2}{2}\!\int\! \frac{\mathrm{d}^Dp}{(2\pi)^D}\frac{ie^{-ip.(x-y)}}{\cos[\theta_{\varepsilon}(\mathbf{p})\delta t]-\cos[p_0\delta t]},
 \end{split}
\end{equation}
where $D=d+1$ and we have $\theta_{\varepsilon}(\mathbf{p})=\theta(\mathbf{p})-i\varepsilon$.  One should take the limit of $\varepsilon\rightarrow 0$ in calculations.  Note that from here on, unless otherwise specified, momentum integrals range over $(-\pi/\delta t,\pi/\delta t]\times(-\pi/a,\pi/a]^d$, where the first factor corresponds to the $p_0$ integral.  In fact, we can also choose our $i\varepsilon$ prescription slightly differently to get
\begin{equation}
\begin{split}
 D_F(x-y) & = \frac{\delta t^2}{2}\!\int\! \frac{\mathrm{d}^Dp}{(2\pi)^D}\frac{ie^{-ip.(x-y)}}{\cos[\theta(\mathbf{p})\delta t]-\cos[p_0\delta t]+i\varepsilon}.
 \end{split}
\end{equation}
In the limit as $a\rightarrow 0$, we recover the usual Feynman propagator.  The momentum-space propagator is
\begin{equation}
\begin{split}
 D_F(p) & = \frac{\delta t^2}{2}\frac{i}{\cos[\theta(\mathbf{p})\delta t]-\cos[p_0\delta t]+i\varepsilon}.
 \end{split}
\end{equation}

\begin{figure}[!ht]
{\centering
\resizebox{7.5cm}{!}{\includegraphics{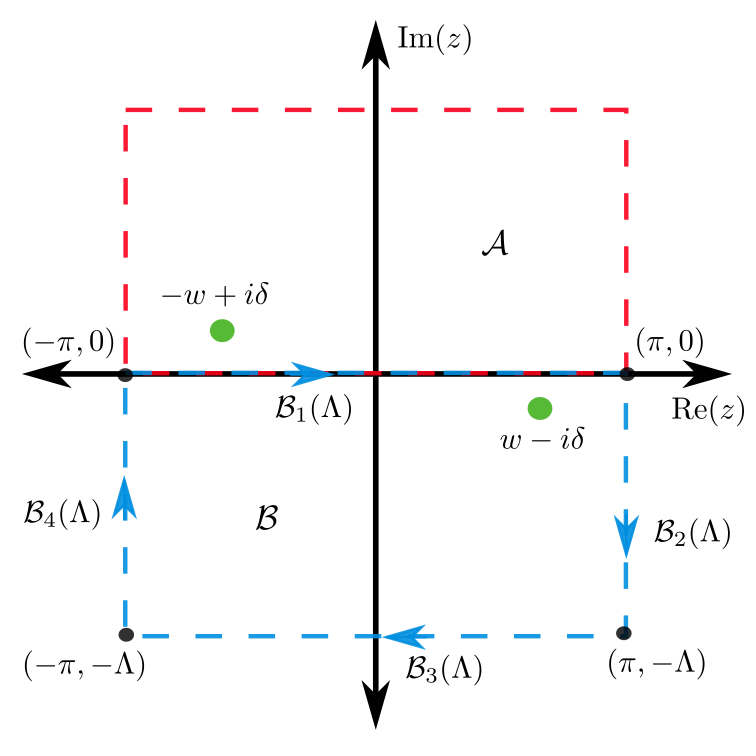}} \caption[Contour]{Contour for the integral in equation (\ref{eq:contour_integral}). \label{fig:contour}}
}
\end{figure}

Let us derive equation (\ref{eq:asdf}).  Substituting the formula for $\phi(x)$ from equation (\ref{eq:phi_again}) into equation (\ref{eq:asdf2}), we get
 \begin{equation}
  D_F(x-y) = \begin{dcases}
                          \int\! \frac{\mathrm{d}^dp}{(2\pi)^d}\frac{e^{-ip_s.(x-y)}}{2\omega(\mathbf{p})} & x^0>y^0\\
                          \int\! \frac{\mathrm{d}^dp}{(2\pi)^d}\frac{e^{ip_s.(x-y)}}{2\omega(\mathbf{p})} & x^0<y^0.
                         \end{dcases}
\end{equation}
We can use the fact that both $\omega(\mathbf{p})=\omega(-\mathbf{p})$ and $\theta(\mathbf{p})=\theta(-\mathbf{p})$ to rewrite this as
\begin{equation}
 \begin{split}
  D_F(x-y) & = \int\! \frac{\mathrm{d}^dp}{(2\pi)^d}\frac{e^{i\mathbf{p}.(\mathbf{x}-\mathbf{y})}e^{-i\theta(\mathbf{p})|x_0-y_0|}}{2\omega(\mathbf{p})} \\
  & = \frac{\delta t}{2}\int\! \frac{\mathrm{d}^dp}{(2\pi)^d}\frac{e^{i\mathbf{p}.(\mathbf{x}-\mathbf{y})}e^{-i\theta(\mathbf{p})|x_0-y_0|}}{\sin[\theta(\mathbf{p})\delta t]},
  \end{split}
\end{equation}
where we used $\omega(\mathbf{p})=\sin[\theta(\mathbf{p})\delta t]/\delta t$ to get the final line.  Note that if $\delta t = a$, this replaced by $\omega(\mathbf{p})=\sin[\theta(\mathbf{p})a]/a$.  To rewrite this in a nicer way, we will show below, via contour integration, that
\begin{equation}\label{eq:tricky}
 \frac{e^{-i\theta_{\varepsilon}(\mathbf{p})|t|}}{\sin[\theta_{\varepsilon}(\mathbf{p})\delta t]} = \delta t\!\int_{-\pi/\delta t}^{\pi/\delta t}\!\frac{\mathrm{d}p_0}{2\pi}\frac{ie^{-ip_0t}}{\cos[\theta_{\varepsilon}(\mathbf{p})\delta t]-\cos(p_0\delta t)},
\end{equation}
where we have $\theta_{\varepsilon}(\mathbf{p})=\theta(\mathbf{p})-i\varepsilon$.  Then we get that the propagator can be written as
\begin{equation}
\begin{split}
 D_F(x-y) & = \frac{\delta t^2}{2}\!\int\! \frac{\mathrm{d}^Dp}{(2\pi)^D}\frac{ie^{-ip.(x-y)}}{\cos[\theta_{\varepsilon}(\mathbf{p})\delta t]-\cos(p_0\delta t)}.
 \end{split}
\end{equation}
So we just need to do the contour integral to get equation (\ref{eq:tricky}).  This is equivalent to proving that
\begin{equation}\label{eq:contour_integral}
\begin{split}
 \frac{e^{-i(w-i\delta)|\tau|}}{\sin(w-i\delta)} & = \int_{-\pi}^{\pi}\frac{\mathrm{d}z}{2\pi}\frac{ie^{-iz\tau}}{\cos(w-i\delta)-\cos(z)}\\
 & = \int_{-\pi}^{\pi}\mathrm{d}zf(z),
 \end{split}
\end{equation}
where we used the substitutions $z=p_0\delta t$, $w=\theta(\mathbf{p})\delta t$, $\delta = \varepsilon \delta t$ and $\tau = t/ \delta t$.  Note that $\tau$ is an integer and $w\in(0,\pi)$.  We can do the integral by contour integration using the contours shown in figure \ref{fig:contour}.  The only poles are located at $z=\pm(w-i\delta)$.  The residues at those points are given by
\begin{equation}
\begin{split}
 \mathrm{Res}(f,w-i\delta) & =\frac{ie^{-i(w-i\delta)\tau}}{2\pi\sin(w-i\delta)}\\
 \mathrm{Res}(f,-w+i\delta) & =-\frac{ie^{i(w-i\delta)\tau}}{2\pi\sin(w-i\delta)}.
 \end{split}
\end{equation}

Let us do the integral for the case with $\tau>0$, which corresponds to the bottom contour $\mathcal{B}$ in figure \ref{fig:contour}.  Then
\begin{equation}
\begin{split}
 \int_{\mathcal{B}(\Lambda)} \mathrm{d}z\,f(z) & =\sum_{i=1}^4\int_{\mathcal{B}_i(\Lambda)} \mathrm{d}z\,f(z)\\
 &  =\int_{\mathcal{B}_1(\Lambda)} \mathrm{d}z\,f(z)+\int_{\mathcal{B}_3(\Lambda)} \mathrm{d}z\,f(z),
 \end{split}
\end{equation}
which follows because $f(\pi+i\lambda)=f(-\pi+i\lambda)$ ensures that the integral over $\mathcal{B}_2(\Lambda)$ cancels that over $\mathcal{B}_4(\Lambda)$.  Then because of the choice of contour, we have that
\begin{equation}
\begin{split}
 & \int_{\mathcal{B}_3(\Lambda)} \mathrm{d}z\,f(z) =\int_{\pi}^{-\pi} \mathrm{d}p\,f(p-i\Lambda)\\
 & = \int_{\pi}^{-\pi}\frac{\mathrm{d}p}{2\pi}\frac{e^{-ip\tau}e^{-\Lambda\tau}}{\cos(w-i\delta)-\cos(p-i\Lambda)},
 \end{split}
\end{equation}
which goes to zero as $\Lambda\rightarrow\infty$ since $\tau>0$.  Then using Cauchy's integral formula, we have that
\begin{equation}
\begin{split}
 \int_{\mathcal{B}_1(\Lambda)} \mathrm{d}z\,f(z) & =(-2\pi i )\frac{ie^{-i(w-i\delta)\tau}}{2\pi\sin(w-i\delta)},\\
 & =\frac{e^{-i(w-i\delta)\tau}}{\sin(w-i\delta)}.
 \end{split}
\end{equation}
where the minus sign in the first line comes from the direction of the contour.  For the integral with $\tau<0$, which corresponds to the top contour $\mathcal{A}$, a similar argument shows that
\begin{equation}
\begin{split}
 \int_{\mathcal{B}_1(\Lambda)} \mathrm{d}z\,f(z) & =\frac{e^{i(w-i\delta)\tau}}{\sin(w-i\delta)}.
 \end{split}
\end{equation}

\section{Interacting fields}
\label{app:int-fields}
In cases where we do not have a correspondence with lattice QFT, it is still possible to include interactions and even do perturbation theory.  In particular, this applies to the Shift circuit, although even in the case of the Strang-split circuit, the discrete-time formalism below can be applied to derive the Feynman rules.  The ideas here may also have other uses in quantum circuit simulations of other field theories when there is no connection to a path integral or to quantum cellular automata more generally.

To include interactions in a free model, we can add further unitaries to the free dynamics:
\begin{equation}
U = U^{1/2}_{\mathrm{int}}U_{0}U^{1/2}_{\mathrm{int}}
\end{equation}
where $U_{0}$ is the unitary evolution operator for the free theory, and $U_{\mathrm{int}}=e^{-iV}$ is the interaction unitary, where $V$ is essentially the analogue of the interaction term in a Hamiltonian.  Note that in the cases we have considered, the unitary is already in this form since we can factor out $U^{1/2}_{\mathrm{int}}=e^{-iV/2}$ with
\begin{equation}
V = a^{d}\delta t\frac{\lambda}{4!}\sum_{\mathbf{n}\in\mathbb{Z}^d}\phi_{\mathbf{n}}^{4}.
\end{equation}
This follows because, in both cases, i.e., equations (\ref{eq:Strang-ver}) and (\ref{eq:shift-dec}), all terms in $W_X$ commute, so $W_X=W^0_Xe^{-iV/2}$.

At this point, it is useful to introduce the interaction picture, where in a rough sense operators evolve via the free dynamics and states evolve via the rest.  More precisely, for operators in the interaction picture, we have
\begin{equation}
 A_I(\tau) = U_{0}^{\dagger \tau} A_S U_{0}^{\tau},
\end{equation}
where $\tau \in \mathbb{Z}$ and $A_S$ denotes an operator in the Schr{\"o}dinger picture.  To find the evolution operator for states, we note that physical quantities cannot depend on which picture we use, so we require that
\begin{equation}
 \langle \psi_f | U^{\dagger \tau} A_S U^{\tau} | \psi_i\rangle = \langle \psi_f | \left[U^{\dagger \tau}U_{0}^{ \tau} \right]A_I(\tau) \left[U_{0}^{\dagger \tau}U^{\tau}\right] | \psi_i\rangle,
\end{equation}
where both $|\psi_i\rangle$ and $|\psi_f\rangle$ are states in the Schr{\"o}dinger picture.  Thus, the evolution operator for states in the interaction picture from time $0$ to time $\tau$ is
\begin{equation}
\begin{split}
 U_{I}(\tau,0) & = U_{0}^{\dagger \tau}U^{\tau}\\
 & = U_{0}^{\dagger \tau}(U^{1/2}_{\mathrm{int}}U_0U^{1/2}_{\mathrm{int}})^{\tau}\\
 & = U_{0}^{\dagger \tau}U^{1/2}_{\mathrm{int}}U_0U^{1/2}_{\mathrm{int}}(U^{1/2}_{\mathrm{int}}U_0U^{1/2}_{\mathrm{int}})^{\tau-1}\\
 & = U^{1/2}_{\mathrm{int},I}(\tau)U^{1/2}_{\mathrm{int},I}(\tau-1)U_{0}^{\dagger \tau-1}(U^{1/2}_{\mathrm{int}}U_0U^{1/2}_{\mathrm{int}})^{\tau-1}\\
 & = U^{1/2}_{\mathrm{int},I}(\tau)U_{\mathrm{int},I}(\tau-1)...U_{\mathrm{int},I}(1)U^{1/2}_{\mathrm{int},I}(0),
 \end{split}
\end{equation}
where $U_{\mathrm{int},I}(\tau)$ is the operator $U_{\mathrm{int}}$ in the interaction picture, i.e., 
\begin{equation}
U_{\mathrm{int},I}(\tau) = U_{0}^{\dagger \tau}U_{\mathrm{int}}U_{0}^{\tau} = e^{-iV_{I}(\tau)}.
\end{equation}
The interaction picture evolution operator for states from time $\tau_1$ to time $\tau_2$ can then be written as
\begin{align}
\begin{split}
 & U_{I}(\tau_{2},\tau_{1}) =\\  &\mathcal{T}\left[\exp\!\left(-i\left[\frac{1}{2}V_{I}(\tau_2)+\sum_{\nu=\tau_{1}-1}^{\tau_{2}-1}V_{I}(\nu)+\frac{1}{2}V_{I}(\tau_1)\!\right]\right)\right].
 \end{split}
\end{align}
This looks the same as the continuum expression, except that instead of an integral over time, we have a sum over discrete times and we have the slightly different boundary conditions at times $\tau_1$ and $\tau_2$.  Since we will take the limit as these go to infinity, this is unimportant.

Note that this is completely general, so we could of course consider different interaction terms like $\phi_{\mathbf{n}}^{3}$.  And the idea applies to other analogous discretizations of QFT where there is some $U_0$ that we can solve and some non-trivial $U_{\mathrm{int}}$.  Also, note that the alternate prescription for adding interactions analogous to the first-order Trotter decomposition $U = U_{0}U_{\mathrm{int}}$ (or $U = U_{\mathrm{int}}U_{0}$)  works in a similar way, but we get the interaction-picture evolution operator
\begin{align}
U_{I}(\tau_{2},\tau_{1}) =\mathcal{T}\left[\exp\!\left(-i\sum_{\nu=\tau_{1}}^{\tau_{2}-1}V_{I}\left(\nu\right)\!\right)\right].
\end{align}
Because Strang splitting is typically more accurate than the first-order Trotter decomposition, one would expect that the symmetric way of including the interaction would be more accurate in practice.  Whichever we choose, the Feynman rules of the next section are the same.  It may be possible to extend these formulas to higher order Suzuki-Trotter decompositions in a similar way, though the expressions would be more complex.

\section{Feynman rules}
\label{sec:Feynman-rules-discrete}
In this section, we will give the Feynman rules for our circuits.  The ideas have the potential to apply to alternative quantum circuits discretizing QFTs that have no obvious connection to Lagrangians or even Hamiltonians.  Getting Feynman rules giving correct descriptions of physical processes from a proposed circuit would then serve as a guide as to whether the proposed circuits are useful discretizations for quantum simulations of a QFT.

The arguments we use to get the Feynman rules are given in section \ref{app:Feyn_rules} and are almost identical to those in continuum QFT (via canonical quantization), which can be found in, e.g., \cite{PeskinSchroeder,Tong06}.  We will not give a rigorous derivation along the lines of an LSZ reduction formula or, e.g., a discrete-time version of the Gell-Mann and Low theorem \cite{51GL}.  Instead, we will start with intuitive notions of asymptotically free states that scatter.  With that in mind, we define the scattering matrix to be
\begin{equation}
S = \lim_{\tau\rightarrow\infty}U_{I}(\tau,-\tau),
\end{equation}
which is analogous to the continuum expression.  The goal is then to approximate scattering amplitudes of the form
\begin{equation}
 \langle f|S| i\rangle,
\end{equation}
where $|i\rangle$ and $|f\rangle$ are initial and final states, e.g., for two incoming particles, we would ideally have
\begin{equation}
|i\rangle = \sqrt{2\omega(\mathbf{p})}\sqrt{2\omega(\mathbf{q})}b_{\mathbf{p}}^{\dagger}b_{\mathbf{q}}^{\dagger}|0\rangle.
\end{equation}
Here are the rules for calculating $\langle f|S| i\rangle$.  A sketch of their derivation is given in section \ref{app:Feyn_rules}.
\begin{enumerate}
    \item Draw all possible amputated and connected diagrams with the right number of incoming and outgoing legs.
	\item Assign a directed momentum to each line, with energy and momentum conservation at each vertex.
	\item Each internal line picks up a momentum-space propagator ${D}_{F}(p)$.
	\item Each vertex gets a factor of $-i\lambda$.
	\item Integrate over all undetermined momenta via
	\begin{equation*}
	 \int\frac{\mathrm{d}^Dp}{(2\pi)^D}.
	\end{equation*}
	\item Divide by the symmetry factor of each diagram.
\end{enumerate}
Note that the symmetry factor is the same as that arising in conventional perturbation theory in QFT for $\phi^4$ theory with the conventional $\lambda/4!$ factor.  The only differences are that the propagator $D_F(p)$ has a different form compared to the continuum, and energy-momentum integrals are over $(-\pi/\delta t,\pi/\delta t]\times(-\pi/a,\pi/a]^d$.  These modifications are not dissimilar to those for the Feynman rules in lattice QFT \cite{Maas2017}.  Indeed, for the Strang-split circuit, these are exactly the same as in discrete-time Lagrangian lattice QFT.  For the Shift circuit, which has no obvious lattice QFT analogue, we have a new propagator.

\section{Deriving the Feynman rules}\label{app:Feyn_rules}
We want to calculate scattering amplitudes like
\begin{equation}
 \langle f|S| i\rangle,
\end{equation}
where $|i\rangle$ and $|f\rangle$ are initial and final states.  For simplicity, let us focus on the case with two incoming and two outgoing particles, so we take
\begin{equation}
\begin{split}
|i\rangle & = |\mathbf{p}_1\mathbf{p}_2\rangle = \sqrt{2\omega(\mathbf{p}_1)}\sqrt{2\omega(\mathbf{p}_2)}b_{\mathbf{p}_1}^{\dagger}b_{\mathbf{p}_2}^{\dagger}|0\rangle\\
|f\rangle & = |\mathbf{k}_1\mathbf{k}_2\rangle = \sqrt{2\omega(\mathbf{k}_1)}\sqrt{2\omega(\mathbf{k}_2)}b_{\mathbf{k}_1}^{\dagger}b_{\mathbf{k}_2}^{\dagger}|0\rangle
.
\end{split}
\end{equation}
Next we insert
\begin{align}
S =\mathcal{T}\left[\exp\!\left(-ia^{d}\delta t\frac{\lambda}{4!}\sum_{x}\phi(x)^{4}\!\right)\right],
\end{align}
where $\sum_x$ is shorthand for the sum over all spacetime points, i.e., $\sum_{\mathbf{n}\in\mathbb{Z}^d}\sum_{\tau\in\mathbb{Z}}$.  Taylor expanding to order $\lambda$, we get
\begin{equation}
 \langle f|S| i\rangle = \langle f| i\rangle-ia^{d}\delta t\frac{\lambda}{4!}\sum_{x}\langle f|\phi(x)^{4}| i\rangle+O(\lambda^2).
\end{equation}
The zeroth order term is simply
\begin{equation}
\begin{split}
\langle f| i\rangle & =  4\omega(\mathbf{p}_1)\omega(\mathbf{p}_2)(2\pi)^6\bigg[\delta^d(\mathbf{p}_1-\mathbf{k}_1)\delta^d(\mathbf{p}_2-\mathbf{k}_2)\\
&+\delta^d(\mathbf{p}_1-\mathbf{k}_2)\delta^d(\mathbf{p}_2-\mathbf{k}_1)\bigg],
\end{split}
\end{equation}
where $\delta^d(\mathbf{p})=\delta_a(p_1)...\delta_a(p_d)$ with $\delta_a(q)$ denoting the lattice momentum delta function with period $2\pi/a$.

The zeroth order term is not so interesting, so we will focus on evaluating $\mathcal{M}$, defined by
\begin{equation}
 \langle f| S-1 |i\rangle = (2\pi)^D\delta^D(p_1+p_2-k_1-k_2)i\mathcal{M},
\end{equation}
where $\delta^D(p)=\delta_{\delta t}(p_0)\delta_a(p_1)...\delta_a(p_d)$.  Here $\delta_a(q)$ denotes the lattice momentum delta function as before, and $\delta_{\delta t}(q)$ denotes the quasi-energy delta function with period $2\pi/\delta t$.  Total energy and momentum will always be conserved, which is the reason we factor this delta function out in our definition of $\mathcal{M}$.

To evaluate the first order term (and the higher order terms), it will help to recall Wick's theorem (see, e.g., \cite{PeskinSchroeder}).  This states that
\begin{equation}
 \mathcal{T}\left[\phi(x_1)...\phi(x_N)\right]=\mathcal{N}\left[\phi(x_1)...\phi(x_N)+\mathrm{all\ contractions}\right].
\end{equation}
Here $\mathcal{N}[\cdot]$ denotes normal ordering of operators, which means that all creation operators are shifted to the left of all annihilation operators, e.g., $\mathcal{N}[b_{\mathbf{p}_1}b^{\dagger}_{\mathbf{p}_2}b_{\mathbf{p}_3}]=b^{\dagger}_{\mathbf{p}_2}b_{\mathbf{p}_1}b_{\mathbf{p}_3}$.  Also, a contraction between two fields is defined to be
\begin{equation}
\wick{\c\phi(x)\c\phi(y)}=\left\langle \mathcal{T}[\phi(x)\phi(y)]\right\rangle =D_F(x-y).
\end{equation}
Then the term ``all contractions'' means the sum of all possible contractions of $\phi(x_1)...\phi(x_N)$, e.g.,
\begin{equation}
\begin{split}
\phi(x_1)\phi(x_2)\phi(x_3)\phi(x_4) & =\wick{\c\phi(x_1)\phi(x_2)\c\phi(x_3)\phi(x_4)}\\
& +\wick{\c1\phi(x_1)\c2\phi(x_2)\c1\phi(x_3)\c2\phi(x_4)}\\
& +...
\end{split}
\end{equation}

\begin{figure}[ht!]
 {\centering
\resizebox{4.5cm}{!}{\includegraphics{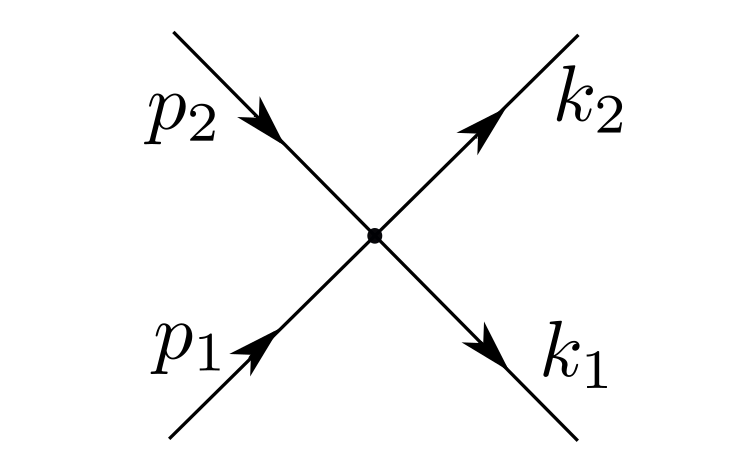}}}
\caption{A diagram contributing to the $O(\lambda)$ component of the scattering amplitude.}
\label{fig:tree}
\end{figure}

Operators that are contracted just contribute a Feynman propagator, whereas uncontracted field operators are cancelled by the creation and annihilation operators of the initial and final states.  So we can define the contraction of a field operator with a creation operator to be
\begin{align}
\wick{\c1 \phi_{I}(x)|\c1{\mathbf{ p}}\rangle=\c2 \phi_{I}(x)\sqrt{2\omega(p)}\c2 b_{\mathbf{p}}^{\dagger}}=e^{-ip.x}.
\end{align}
Then to calculate, e.g., the $O(\lambda)$ term, we evaluate all possible contractions of the fields and the initial and final state creation and annihilation operators.  As an example contributing to the $O(\lambda)$ term, consider the contraction
\begin{equation}
\begin{split}
\wick{\langle\c1{\mathbf{k}_1}\c2{\mathbf{k}_2}|\c1\phi(x)\c2\phi(x)}\wick{\c1\phi(x)\c2\phi(x)|\c1{\mathbf{p}_1}\c2{\mathbf{p}_2}\rangle} = e^{-i(p_1+p_2-k_1-k_2).x}.
 \end{split}
\end{equation}
Then the overall contribution to $i\mathcal{M}$ from this contraction is
\begin{equation}
\begin{split}
 &-ia^{d}\delta t\frac{\lambda}{4!}\sum_{x}e^{-i(p_1+p_2-k_1-k_2).x}\\
 &=-i\frac{\lambda}{4!}(2\pi)^D\delta^D(p_1+p_2-k_1-k_2),
 \end{split}
\end{equation}
where we used the identities $\sum_{n\in\mathbb{Z}}e^{inz}=2\pi\delta(z)$ and $\delta(Cz)=\delta(z)/C$.  We see the factors of $(2\pi)^D\delta^D(p_1+p_2-k_1-k_2)$ as expected from our definition of $\mathcal{M}$.  We assign the diagram in figure \ref{fig:tree} to this contraction.

In fact, there are $4!$ possible contractions giving rise to the same diagram, which cancels the $1/4!$ factor in the interaction and simplifies the Feynman rules.  The $1/4!$ factors do not always cancel for a given diagram, which is the reason for symmetry factors in the Feynman rules, as we will see.

\begin{figure}[ht!]
 {\centering
\resizebox{4.5cm}{!}{\includegraphics{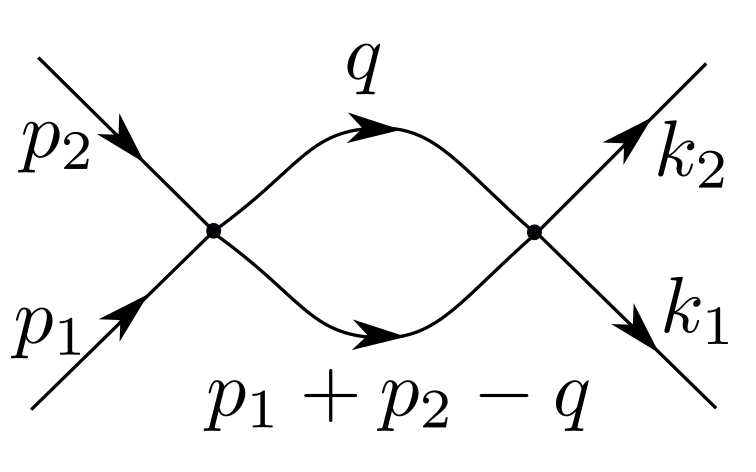}}}
\caption{A diagram contributing to the $O(\lambda^2)$ component of the scattering amplitude.}
\label{fig:loop}
\end{figure}

A more interesting example comes from the order $\lambda^2$ contribution to the amplitude.  Consider
\begin{equation}
\begin{split}
& \wick{\langle\c1{\mathbf{k}_1}\c2{\mathbf{k}_2}|\c1\phi(x)\c2\phi(x)}\wick{\c1\phi(x)\c2\phi(x)\c1\phi(y)\c2\phi(y)}\wick{\c1\phi(y)\c2\phi(y)|\c1{\mathbf{p}_1}\c2{\mathbf{p}_2}\rangle}\\
 & = e^{-i(p_1+p_2).x+i(k_1+k_2).y}D_F(x-y)D_F(x-y).
 \end{split}
\end{equation}
The full contribution  to the amplitude is then
\begin{equation}
\begin{split}
 &-a^{2d}\delta t^2\frac{\lambda^2}{2(4!)^2}\sum_{x,y}e^{-i(p_1+p_2).x+i(k_1+k_2).y}D_F(x-y)^2\\
 &=\frac{-\lambda^2}{2(4!)^2}(2\pi)^{2D}\!\int\!\frac{\mathrm{d}^Dq_1\mathrm{d}^Dq_2}{(2\pi)^{2D}}\delta^D(p_1+p_2-q_1-q_2)\times\\
 &\ \ \ \times \delta^D(q_1+q_2-k_1-k_2)D_F(q_1)D_F(q_2)\\
 &=\frac{-\lambda^2}{2(4!)^2}(2\pi)^{D}\delta^D(p_1+p_2-k_1-k_2)\times\\
 &\ \ \ \times\!\int\!\frac{\mathrm{d}^Dq}{(2\pi)^{D}} D_F(q)D_F(p_1+p_2-q),
 \end{split}
\end{equation}
where the second line followed by plugging in the definition of the Feynman propagator:
\begin{equation}
\begin{split}
 D_F(x-y) & = \!\int\! \frac{\mathrm{d}^Dp}{(2\pi)^D}e^{-ip.(x-y)}D_F(p)\\
 & = \!\int\! \frac{\mathrm{d}^Dp}{(2\pi)^D}e^{ip.(x-y)}D_F(p),
 \end{split}
\end{equation}
where by abuse of notation $D_F(p)$ denotes the momentum space propagator, and we used that $D_F(p)=D_F(-p)$.  The  diagram corresponding to this process is given in figure \ref{fig:loop}.
The number of contractions contributing to this diagram is $(4!)^2$, which cancels the $1/(4!)^2$ factor and means that the symmetry factor is $1/2$, since a factor of $1/2$ came from the Taylor expansion of $S$.  In practice, one finds the symmetry factor by looking at the symmetries of the diagram using the rules given in, e.g., \cite{PeskinSchroeder}, as opposed to counting the number of different contractions.  The important point for us is that the symmetry factors for diagrams are the same as those in QFT.

\begin{figure}[ht!]
 {\centering
\resizebox{4.5cm}{!}{\includegraphics{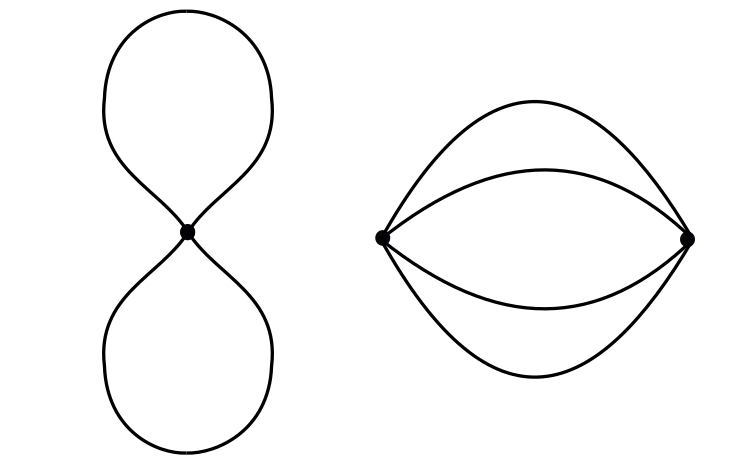}}}
\caption{A diagram describing vacuum bubbles that do not contribute to scattering amplitudes.}
\label{fig:vacdiag}
\end{figure}

This procedure of contracting fields by hand for terms in the expansion of $S$ can be replaced by the Feynman rules. These rules (in momentum space) for calculating $i\mathcal{M}$ are as follows.
\begin{enumerate}
    \item Draw all possible amputated and connected diagrams with the right number of incoming and outgoing legs.
	\item Assign a directed momentum to each line, with energy and momentum conservation at each vertex.
	\item Each internal line picks up a momentum-space propagator ${D}_{F}(p)$.
	\item Each vertex gets a factor of $-i\lambda$.
	\item Integrate over all undetermined momenta via
	\begin{equation*}
	 \int\frac{\mathrm{d}^Dp}{(2\pi)^D}.
	\end{equation*}
	\item Divide by the symmetry factor of each diagram.
\end{enumerate}
We still have to justify the assumption of ``amputated and connected'' diagrams.  First, ``connected'' means that every part of the diagram is connected to at least one external leg.  Basically, this excludes vacuum bubbles, meaning it rules out diagrams such as that in figure \ref{fig:vacdiag}.
One way to motivate this is to argue that the following formula, relating correlation functions in the interacting theory to calculations with the free vacuum, should hold:
\begin{equation}\label{eq:corr-S}
 \langle \Omega|\phi_H(x_1)...\phi_H(x_N)|\Omega \rangle = \frac{\langle 0|\mathcal{T}\left[\phi(x_1)...\phi(x_N)S\right]|0 \rangle}{\langle 0|S|0 \rangle},
\end{equation}
where $|\Omega\rangle$ is a state analogous to the vacuum in the presence of interactions and $\phi_H(x)=U^{-x_0/\delta t}\phi_{\mathbf{x}/a}U^{x_0/\delta t}$ denotes the field in the Heisenberg picture.  The division by $\langle 0|S|0 \rangle$ on the right hand side is what cancels vacuum bubbles and justifies the restriction to connected diagrams.  

Equation (\ref{eq:corr-S}) holds if we can make sense of
\begin{equation}\label{eq:vac}
 |\Omega\rangle \langle \Omega |0\rangle = \lim_{T\rightarrow \infty}U^{T}|0\rangle.
\end{equation}
To see this, we assume for simplicity that $x^0_1\geq x^0_2\geq...\geq x^0_N$ to get
\begin{equation}
\begin{split}
& \langle 0|\mathcal{T}\left[\phi(x_1)...\phi(x_N)S\right]|0 \rangle\\
& = \lim_{T\rightarrow \infty}\langle 0|U(T,t_1)\phi(x_1)U(t_1,t_2)\phi(x_2)...U(t_N,-T)|0 \rangle\\
& = \lim_{T\rightarrow \infty}\langle 0|U^T\phi_H(x_1)...\phi_H(x_N)U^T|0 \rangle.
\end{split}
\end{equation}
To get the second line, we use that the time-ordering operator splits up $S=\lim_{T\rightarrow \infty}U(T,-T)$ into each of the $U(t_i,t_j)$ terms, where we are abusing the notation a little to identify $U(t_i,t_j)=U(\tau_i,\tau_j)$, with $t_i=\tau_i\delta t$.  The last line follows by using the definition of the field operators in the interaction picture, $\phi(x)=U_0^{\dagger x_0/t}\phi_{\mathbf{x}/a}U_0^{x_0/\delta t}$, and the definition of $U(t,0)=U(\tau,0)=U_0^{\dagger\tau}U^{\tau}$.  It may be useful to note that $U(t_1,t_2)=U(t_1,0)U(t_2,0)^{-1}$.

We can give a rough justification for equation (\ref{eq:vac}), if the operator $U$ has an eigenvector $|\Omega\rangle$ with eigenvalue $1$ and the rest of its spectrum is continuous.  Then, following \cite{Tong06}, we can (very roughly) argue that, for any state $|\psi\rangle$,
\begin{equation}\label{eq:limOm}
\begin{split}
 & \lim_{T\rightarrow \infty}\langle \psi|U^{T}|0\rangle\\
 & =\langle\psi|\Omega\rangle \langle \Omega |0\rangle+\lim_{T\rightarrow \infty}\int\mathrm{d}\alpha e^{-i\alpha T}\langle \psi|\alpha\rangle \langle \alpha |0\rangle\\
 & =\langle\psi|\Omega\rangle \langle \Omega |0\rangle,
 \end{split}
\end{equation}
where $U|\alpha\rangle = e^{-i\alpha}|\alpha\rangle$ and $\alpha$ labels the continuous part of the spectrum.
Getting the last line of equation (\ref{eq:limOm}) is justified by the Riemann-Lebesgue lemma, which tells us that, for an $L^1$ function $f(x)$, $\int_{\mathbb{R}}\mathrm{d}x f(x)e^{-ixz}\rightarrow 0$ as $|z|\rightarrow \infty$.  In this case, we are assuming that $\langle \psi|\alpha\rangle \langle \alpha |0\rangle$ is an $L^1$ function of $\alpha$.  This is not a rigorous argument, which instead we postpone to future work.

Regarding ``amputated'' diagrams, this means that we do not include any diagrams with loops on external legs.  Loops on external legs correspond to the fact that the propagator of the in-going and out-coming particles is modified in the interacting theory, which is motivated properly via the LSZ reduction theorem \cite{PeskinSchroeder}, which describes how to correctly relate $S$ matrix elements to correlation functions.  Again we postpone a thorough investigation of this to future work.
    
\section{One-loop calculation}\label{app:One_loop}
\begin{figure}[ht!]{\centering
\resizebox{3.0cm}{!}{\includegraphics{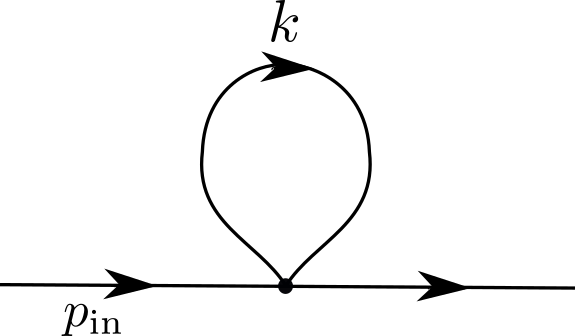}}} 
\caption{One-loop correction to the propagator in the presence of interactions.}
\label{fig:one--loop}
\end{figure}

We want to evaluate the one-loop correction to the mass for different regulators.  This corresponds to the diagram in figure \ref{fig:one--loop}.
Let us consider the case of continuum QFT with a momentum cutoff $\Lambda$, which is given by
 \begin{equation}
 \begin{split}
 \Pi_{\mathrm{cont}}(p_{\mathrm{in}}^2) & = -i\lambda\frac{1}{2}\!\int\!\! \frac{\mathrm{d}^2p}{(2\pi)^2}D^{\mathrm{cont}}_F(p) \\
 & = \frac{\lambda}{8\pi}\!\int_{-\Lambda}^{\Lambda}\! \frac{\mathrm{d}p}{\sqrt{p^2+m^2}} \\
 & = \frac{\lambda}{4\pi}\ln\left(\frac{\Lambda}{m}\right)+\mathrm{finite},
 \end{split}
 \end{equation}
 where $D^{\mathrm{cont}}_F(p)=i/(p^2+m^2)$ is the continuum propagator in momentum space, and the factor of $1/2$ in the first line comes from the symmetry factor.  We see logarithmic divergence with the cutoff.

Next, we consider the Shift circuit.  Using the Feynman rules, with the factor $-i\lambda$ assigned to each vertex, we have to evaluate
 \begin{equation}
\begin{split}
  \Pi_{\mathrm{shift}}(p_{\mathrm{in}}^2) & =-i\lambda\frac{1}{2}\!\int\! \frac{\mathrm{d}^2p}{(2\pi)^2}D_F(p)\\
  & =-i\lambda\frac{1}{2}\!\int\! \frac{\mathrm{d}^2p}{(2\pi)^2}\frac{a^2}{2}\frac{i}{M\cos(p_1a)-\cos(p_0a)+i\varepsilon}\\
  & =\frac{\lambda}{4}\!\int\! \frac{\mathrm{d}p}{2\pi}\frac{a}{\sqrt{1-(M\cos(pa))^2}},
  \end{split}
 \end{equation}
 where $M=1-m^2a^2/2$.  We can evaluate this integral in terms of the complete elliptic integral of the first kind:
  \begin{equation}
\begin{split}
  \Pi_{\mathrm{shift}}(p_{\mathrm{in}}^2) & =\frac{\lambda}{4}\!\int_{-\pi/a}^{\pi/a}\! \frac{\mathrm{d}p}{2\pi}\frac{a}{\sqrt{1-M^2\cos(pa)^2}}\\
  & =\frac{\lambda}{8\pi}\!\int_{-\pi}^{\pi}\! \mathrm{d}\theta \frac{1}{\sqrt{1-M^2\cos(\theta)^2}}\\
   & =\frac{\lambda}{2\pi}\!\int_{0}^{\pi/2}\! \mathrm{d}\theta \frac{1}{\sqrt{1-M^2\sin(\theta)^2}}\\
   & =\frac{\lambda}{2\pi}K(M),
 \end{split}
 \end{equation}
 where $K(x)$ is the complete elliptic integral of the first kind.  This can be expanded for $|x|$ close to one as \cite{mathworld-elliptic}
 \begin{equation}
  K(x)=-\frac{1}{2}\ln(1-x^2)+\mathrm{const}.
 \end{equation}
Note that in the notation in \cite{mathworld-elliptic}, $K(x)$ is actually denoted by $K(x^2)$.  Plugging this in, we get
  \begin{equation}
 \begin{split}
 \Pi_{\mathrm{shift}}(p_{\mathrm{in}}^2) & =-\frac{\lambda}{4\pi}\ln(1-M^2)+\mathrm{finite}\\
 & =-\frac{\lambda}{4\pi}\ln[m^2a^2+O(a^4)]+\mathrm{finite}\\
 & =\frac{\lambda}{2\pi}\ln\left(\frac{1}{ma}\right)+\mathrm{finite}.
 \end{split}
 \end{equation}
 So we get a prefactor that is off from the momentum cuttoff regulator by a factor of two.  This can be traced back to the high-momentum modes with small values of $\omega(p)=\sqrt{1-M^2\cos(pa)^2}/a$ discussed in section \ref{sec:mainsol}.  To remedy this, let us try instead using a modified interaction term:
 \begin{equation}
V = a^{D}\frac{\lambda}{4!}\sum_{\mathbf{n}\in\mathbb{Z}^d}\left(\sum_{\mathbf{e}\in \mathcal{K}}w(\mathbf{e})\phi_{\mathbf{n}+\mathbf{e}}\right)^{4},
 \end{equation}
 where $\mathcal{K}$ is the set of all vectors with components in $\{-1,0,1\}$ and $w(\mathbf{e})=v(e_1)\times...\times v(e_d)$ with $v(-1)=v(1)=1/4$ and $v(0)=1/2$.  The weights $w(\mathbf{e})$ were chosen in order to have
 \begin{equation}
  \sum_{\mathbf{e}\in \mathcal{K}}w(\mathbf{e})e^{i\mathbf{p}.\mathbf{e}a}=\prod_{i=1}^d\frac{1+\cos[p_i a]}{2}.
 \end{equation}
 As a result of this, the Feynman rules are modified, with the sole change being that each vertex gets a factor of
\begin{equation}
 -i\lambda\prod_{\chi=1}^4\left[ \prod_{i=1}^d\frac{1+\cos[p^{\chi}_i a]}{2}\right],
\end{equation}
where $p^{\chi}_i$ is the $i$th component of the momentum of the $\chi$th line joining the vertex.
 
\begin{figure}[!ht]
{\centering
\resizebox{8.6cm}{!}{\includegraphics{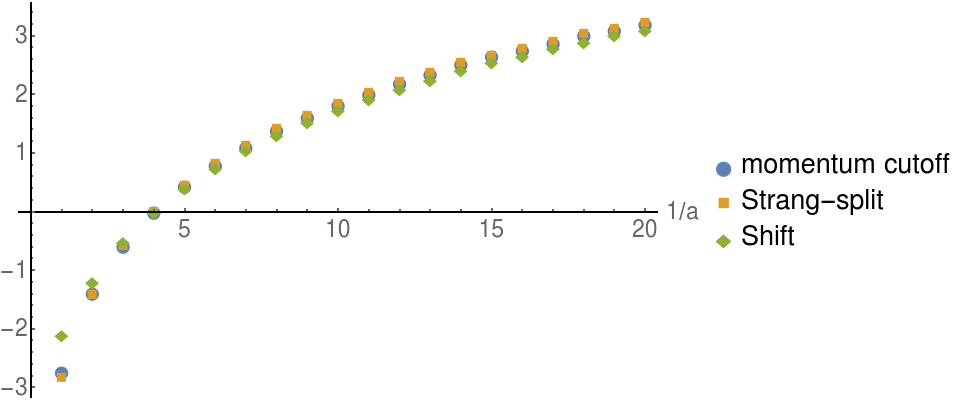}} \caption[Contour]{Here we see the corrections to mass ($\Pi_{\mathrm{cont}}(p_{\mathrm{in}}^2)$ and $\Pi_{\mathrm{shift}}(p_{\mathrm{in}}^2)$) integrated numerically and plotted as a function of the lattice spacing $a$.  Each function has been normalized by subtracting a constant so that they all agree for $a=2$.  We see logarithmic growth with the cutoff in each case with the same prefactor.  For the case of a momentum cutoff, we take $\Lambda = \pi/a$.  The $x$ axis corresponds to $1/a$, i.e., the inverse lattice spacing, and the $y$ axis is in arbitrary units.  In this case, we have taken $p_{\mathrm{in}}=0$.  \label{fig:Mass_correction}}
}
\end{figure}

Now, the new integral for the one-loop diagram is
\begin{equation}\label{eq:one-loop-new-int}
\begin{split}
  & \Pi_{\mathrm{shift}}(p_{\mathrm{in}}^2)\\
  & =\frac{\lambda}{4}\frac{(1+\cos[p_{\mathrm{in}} a])^2}{16}\!\int\! \frac{\mathrm{d}p}{2\pi}\frac{a[1+\cos(p a)]^2}{\sqrt{1-M^2\cos(pa)^2}}.
  \end{split}
 \end{equation}
 Let us ignore the prefactor and focus on the integral for now:
 \begin{equation}
\begin{split}
 & \!\int_{-\pi/a}^{\pi/a}\! \frac{\mathrm{d}p}{2\pi}\frac{a[1+\cos(p a)]^2}{\sqrt{1-M^2\cos(pa)^2}}\\
 & = \!\int_{-\pi}^{\pi}\! \frac{\mathrm{d}\theta}{2\pi}\frac{[1+\cos(\theta)]^2}{\sqrt{1-M^2\cos(\theta)^2}}\\
 & = \!\int_{-\pi}^{\pi}\! \frac{\mathrm{d}\theta}{2\pi}\frac{1+\cos(\theta)^2}{\sqrt{1-M^2\cos(\theta)^2}}.
  \end{split}
 \end{equation}
 The last line follows because the integral vanishes for the terms with numerator linear in $\cos(\theta)$.  To see this, use $\cos(x)=-\cos(x+\pi)$.  Then we get
 \begin{equation}
\begin{split}
 & \!\int_{-\pi}^{\pi}\! \frac{\mathrm{d}\theta}{2\pi}\left[\frac{1+1/M^2}{\sqrt{1-M^2\cos(\theta)^2}} +\frac{M^2\cos(\theta)^2-1}{M^2\sqrt{1-M^2\cos(\theta)^2}}\right]\\
 & = \frac{2(1+1/M^2)K(M)}{\pi}- \!\int_{-\pi}^{\pi}\! \frac{\mathrm{d}\theta}{2\pi} \frac{\sqrt{1-M^2\cos(\theta)^2}}{M^2}\\
 & = \frac{2(1+1/M^2)K(M)}{\pi}+\mathrm{finite},
  \end{split}
 \end{equation}
 where the last line follows from $0\leq\sqrt{1-M^2\cos(\theta)^2}\leq 1$ and $M=1-m^2a^2/2$, so $1/M^2=1+O(a^2)$ for small $a$.
 Then we get 
 \begin{equation}
 \begin{split}
  \!\int_{-\pi/a}^{\pi/a}\! \frac{\mathrm{d}p}{2\pi}\frac{a[1+\cos(p a)]^2}{\sqrt{1-M^2\cos(pa)^2}} & = \frac{4K(M)}{\pi} + \mathrm{finite}.
  \end{split}
 \end{equation}
 This gives us
\begin{equation}
\begin{split}
  & \Pi_{\mathrm{shift}}(p_{\mathrm{in}}^2) =\frac{(1+\cos[p_{\mathrm{in}} a])^2}{16}\frac{\lambda K(M)}{\pi} + \mathrm{finite}.
  \end{split}
 \end{equation}
 Using the same expansion for the elliptic integral $K$ as before, we get finally
 \begin{equation}
\begin{split}
  & \Pi_{\mathrm{shift}}(p_{\mathrm{in}}^2) =\frac{(1+\cos[p_{\mathrm{in}} a])^2}{4}\frac{\lambda}{4\pi}\ln\left(\frac{1}{ma}\right) + \mathrm{finite}.
  \end{split}
 \end{equation}
 For small incoming momenta $p_{\mathrm{in}}$ compared to $1/a$, this agrees well with the calculation earlier of $\Pi_{\mathrm{cont}}(p_{\mathrm{in}}^2)$ since $(1+\cos[p_{\mathrm{in}} a])^2=4+O(a^2)$.  For comparison, the functions $\Pi_{\mathrm{cont}}(p_{\mathrm{in}}^2)$ and $\Pi_{\mathrm{shift}}(p_{\mathrm{in}}^2)$ integrated numerically are plotted in figure \ref{fig:Mass_correction} as a function of the lattice spacing $a$, where we take the momentum cutoff $\Lambda = \pi/a$ for comparison.  Note that in the case of $\Pi_{\mathrm{shift}}(p_{\mathrm{in}}^2)$, we are considering the modified interaction term, so we are integrating equation (\ref{eq:one-loop-new-int}) numerically.
 
\section{Non-abelian circuit and path integral}
\label{app:non-ab-circ}
For a good description of how to connect the transfer matrix to the path integral for lattice gauge theory, see \cite{Smit02}.  Here we allow $\delta t/a =\kappa$, which may or may not equal one.

Again, we consider a basis describing the field configurations on the lattice at time $\eta$ such that
\begin{equation}
 \hat{U}^{ab}_{\mathbf{x},\mathbf{e}}|U(\eta)\rangle = U^{ab}_{\mathbf{x},\mathbf{e}}(\eta)|U(\eta)\rangle,
\end{equation}
We have the initial and final field configurations $|U(0)\rangle$ and $|U(\tau)\rangle$ at time $0$ and time $\tau$ respectively.  The amplitude for one to evolve into the other is
\begin{equation}
 \langle U(\tau)|\hat{T}^{\tau}|U(0)\rangle = \langle U(\tau)|(\hat{W}_{\mathrm{el}}\hat{W}_{\mathrm{mag}})^{\tau}|U(0)\rangle,
\end{equation}
where we have
\begin{equation}
 \begin{split}
  \hat{W}_{\mathrm{mag}} & =\exp\left(-i\frac{2\kappa}{g^2}\sum_{\mathbf{p}_s}\mathrm{Re}\left(\mathrm{tr}[\hat{U}_{\mathbf{p}_s}]\right)\right)\\
  \hat{W}_{\mathrm{el}} & =\prod_{\mathbf{x},\mathbf{e}}\int \mathrm{d}V_{\mathbf{x},\mathbf{e}} \exp\left(-i\frac{2}{\kappa g^2}\mathrm{Re}\left(\mathrm{tr}[V_{\mathbf{x},\mathbf{e}}]\right)\right)\hat{L}_{\mathbf{x},\mathbf{e}}(V_{\mathbf{x},\mathbf{e}}).
 \end{split}
\end{equation}
Inserting factors of $\openone=\int\mathrm{d}U(\eta)|U(\eta)\rangle\langle U(\eta)|$, we get
\begin{equation}
 \langle U(\tau)|\hat{T}^{\tau}|U(0)\rangle = \int\prod_{1}^{\tau-1}\mathrm{d}U(\eta)\prod_{\nu=0}^{\tau-1}\langle U(\nu+1)|\hat{T}|U(\nu)\rangle.
\end{equation}
Dropping the explicit time dependence, we can simplify this via
\begin{equation}
\langle U^{\prime}|\hat{W}_{\mathrm{el}}\hat{W}_{\mathrm{mag}}|U\rangle = \langle U^{\prime}|\hat{W}_{\mathrm{el}}|U\rangle e^{-i\frac{2\kappa}{g^2}\sum_{\mathbf{p}_s}\mathrm{Re}(\mathrm{tr}[U_{\mathbf{p}_s}])}.
\end{equation}
Dealing with $\hat{W}_{\mathrm{el}}$ is more tricky.  First notice that
\begin{equation}
\begin{split}
 & \langle U^{\prime}|\hat{W}_{\mathrm{el}}|U\rangle\\
 & = \langle U^{\prime}|\prod_{\mathbf{x},\mathbf{e}}\int \mathrm{d}V_{\mathbf{x},\mathbf{e}} \exp\left(-i\frac{2}{\kappa g^2}\mathrm{Re}\left(\mathrm{tr}[V_{\mathbf{x},\mathbf{e}}]\right)\right)\hat{L}_{\mathbf{x},\mathbf{e}}(V_{\mathbf{x},\mathbf{e}})|U\rangle\\
 & = \prod_{\mathbf{x},\mathbf{e}}\exp\left(-i\frac{2}{\kappa g^2}\mathrm{Re}(\mathrm{tr}[U_{\mathbf{x},\mathbf{e}}U^{\prime \dagger}_{\mathbf{x},\mathbf{e}}])\right).
 \end{split}
\end{equation}
At this point, we need to use Gauss' law, which tells us that states are invariant under the transformation $U\rightarrow U^{\Omega}$, defined by
\begin{equation}
 U_{\mathbf{x},\mathbf{e}}\rightarrow \Omega_{\mathbf{x}}U_{\mathbf{x},\mathbf{e}}\Omega^{\dagger}_{\mathbf{x}+\mathbf{e}} = U^{\Omega}_{\mathbf{x},\mathbf{e}}
\end{equation}
for each $\mathbf{x}$ and $\mathbf{e}$.  Denoting the unitary operator that implements this gauge transformation on states by $D(\Omega)$, we get
\begin{equation}
 D(\Omega)|U\rangle = | U^{\Omega}\rangle = | U\rangle,
\end{equation}
where the last equality follows from Gauss' law.  Averaging over $\Omega$, we get
\begin{equation}
\begin{split}
 &\langle U^{\prime}|\hat{W}_{\mathrm{el}}|U\rangle = \int\prod_{\mathbf{y}}\mathrm{d}\Omega_{\mathbf{y}}\langle U^{\prime}|\hat{W}_{\mathrm{el}}D(\Omega)|U\rangle \\
 & = \int\prod_{\mathbf{y}}\mathrm{d}\Omega_{\mathbf{y}}\prod_{\mathbf{x},\mathbf{e}}\exp\left(-i\frac{2}{\kappa g^2}\mathrm{Re}(\mathrm{tr}[\Omega_{\mathbf{x}} U_{\mathbf{x},\mathbf{e}}\Omega^{\dagger}_{\mathbf{x}+\mathbf{e}}U^{\prime \dagger}_{\mathbf{x},\mathbf{e}}])\right).
 \end{split}
\end{equation}
If we put back in the time dependence, we can write $U_{\mathbf{x},\mathbf{e}}(\eta)=U_{x,\mathbf{e}}$, where $x=(\eta\delta t,\mathbf{x})$ labels spacetime points.  Then, since we are integrating over gauge transformations for each time $\eta$, we can just as well rename $\Omega_{\mathrm{x}}(\eta) = U_{x,e^0}$, where $e^0$ denotes a lattice basis vector in the positive time direction, to get
\begin{equation}
\begin{split}
 &\langle U^{\prime}|\hat{W}_{\mathrm{el}}|U\rangle \\
 & = \int\prod_{\mathbf{y}}\mathrm{d}U_{y,e^0}\prod_{\mathbf{x},\mathbf{e}}\exp\left(-i\frac{2}{\kappa g^2}\mathrm{Re}(\mathrm{tr}[U_{x,e^0} U_{x,\mathbf{e}}U^{\dagger}_{x+\mathbf{e},e^0}U^{\prime \dagger}_{x,\mathbf{e}}])\right).
 \end{split}
\end{equation}
But $\mathrm{tr}[U_{x,e^0} U_{x,\mathbf{e}}U^{\dagger}_{x+\mathbf{e},e^0}U^{\prime \dagger}_{x,\mathbf{e}}]$ is just the trace of a plaquette unitary, where one of the directions is along the time axis.  We can denote all such plaquette unitary by $U_{p_t}$.  Then, denoting spatial plaquettes by $p_s$, we get
\begin{equation}
 \langle U_{\mathrm{f}}|U^{\tau}|U_{\mathrm{i}}\rangle = \int \mathcal{D}(U)e^{iS(U)},
\end{equation}
where the action $S(U)$ is given by
\begin{equation}
S(U) = \frac{2}{g^2}\left(\frac{a}{\delta t}\sum_{p_t}\mathrm{Re}\left(\mathrm{tr}[U_{p_t}]\right)+\frac{\delta t}{a}\sum_{p_s}\mathrm{Re}\left(\mathrm{tr}[U_{p_s}]\right)\right),
\end{equation}
where the sum over spatial plaquettes does not include those at time $\tau$.
In the case where $\delta t = a$, this simplifies to become
\begin{equation}
S(U) = \frac{2}{g^2}\sum_{p}\mathrm{Re}\left(\mathrm{tr}[U_{p}]\right),
\end{equation}
where the sum is over all spacetime plaquette unitaries in $\{0,...,\tau\}\times \mathbb{Z}^d$ except for spatial plaquettes at time $\tau$.
And the measure is given by
\begin{equation}
 \mathcal{D}(U)=\prod_{x,e}\mathrm{d}U_{x,e},
\end{equation}
where the product includes all spatial links $(x,e)$ with $x^0\in\{1,...,\tau-1\}$ and all temporal links $(x,e)$ with $x^0\in\{0,...,\tau-1\}$.  Note that this path integral formalism is mathematically well defined on finite lattices, at least for compact gauge groups.

\end{document}